\documentclass[reprint,superscriptaddress,amsmath,amssymb,aps,pre,floatfix]{revtex4-1}
\usepackage{graphicx }
\usepackage{dcolumn} 
\usepackage{bm} 
\usepackage{dutchcal} 
\usepackage{color}
\usepackage{bbold}
\usepackage{soul}
\usepackage{amsmath,amssymb}
\newcommand{\newc}{\newcommand}
\newc{\beq}{\begin{equation}}
\newc{\eeq}{\end{equation}}
\newc{\kt}{\rangle}
\newc{\br}{\langle}
\newc{\beqa}{\begin{eqnarray}}
\newc{\eeqa}{\end{eqnarray}}
\newc{\longra}{\longrightarrow}

\begin{document}

\title{Complex saddle trajectories for multidimensional quantum wave packet/coherent state propagation: application to a many-body system}

\author{Steven Tomsovic}
\affiliation{Institut f\"ur Theoretische Physik, Universit\"at Regensburg, 93040 Regensburg, Germany}
\affiliation{Department of Physics and Astronomy, Washington State University, Pullman, WA 99164-2814}

\date{\today}

\begin{abstract}

A practical search technique for finding the complex saddle points used in wave packet or coherent state propagation is developed which works for a large class of Hamiltonian dynamical systems with many degrees of freedom.  The method can be applied to problems in atomic, molecular, and optical physics, and other domains.   A Bose-Hubbard model is used to illustrate the application to a many-body system where discrete symmetries play an important and fascinating role.  For multidimensional wave packet propagation, locating the necessary saddles involves the seemingly insurmountable difficulty of solving a boundary value problem in a high-dimensional complex space, followed by determining whether each particular saddle found actually contributes.  In principle, this must be done for each propagation time considered.  The method derived here identifies a real search space of minimal dimension, which leads to a complete set of contributing saddles up to intermediate times much longer than the Ehrenfest time scale for the system.  The analysis also gives a powerful tool for rapidly identifying the various dynamical regimes of the system.

\end{abstract}

\pacs{}

\maketitle

There are numerous motivations for studying the propagation of multidimensional Gaussian wave packets in quantum systems.  Particularly noteworthy are their intimate connections to coherent states and thus many-body bosonic systems~\cite{Glauber63}, coherent state representations of path integrals~\cite{Baranger01}, molecular spectroscopy~\cite{Heller81b, Gruebele92}, femto-chemistry~\cite{Zewail00}, attosecond physics~\cite{Agostini04}, far out-of-equilibrium dynamics in bosonic many-body systems~\cite{Greiner02b,Polkovnikov11,Tomsovic18b}, and studies of the quantum-classical correspondence and Ehrenfest time scales~\cite{Berry79b,Tomsovic93}.  The Ehrenfest time scale plays a particularly significant role in scrambling~\cite{Sekino08} and many-body quantum interference properties~\cite{Tomsovic18b}.  

In principle, in system regimes alluded to above, precisely where quantum dynamics methods may be effectively impossible to carry out, a time-dependent semiclassical theory can provide both a very accurate approximation to wave packet propagation up to intermediate time scales~\cite{Sepulveda92}, and a very physical interpretation of the results.  Indeed, examples of `simple' chaotic dynamical systems, which have their own challenges, were treated successfully roughly twenty-five years ago~\cite{Oconnor92, Tomsovic91b, Tomsovic93b}.  The theoretical foundations for such wave packet dynamics have existed even longer under various guises~\cite{Littlejohn86}, i.e.~semiclassical approximations to coherent state Feynman path integrals~\cite{Klauder78}, generalized Gaussian wave packet dynamics (GGWPD)~\cite{Huber88}, or a complexification of Maslov's version of time-dependent WKB theory~\cite{Maslov81}.  

Nevertheless, applications to physical systems possessing more than a couple degrees of freedom are largely absent, and extensions must be developed in order to apply these methods in the domain of many-body physics; see for example~\cite{Engl14b,Engl14,Engl15}.  In the Maslov version of time-dependent WKB~\cite{Maslov81}, an initial state is associated with a Lagrangian manifold of phase points, which must be propagated classically for a time $t$, and then intersected with another manifold associated with a final state.  The intersection points identify trajectories, whose initial conditions are points on the initial manifold that propagate to points on the final manifold.  They are the stationary phase points upon which the theory rests.  Numerically solving for these trajectories is tantamount to solving a boundary value problem, which rapidly becomes prohibitive to solve as the number of degrees of freedom increases.  

Confronted with this situation, many studies, particularly in the domain of molecular spectroscopy, have focussed on working around its solution.  For example, a number of methods convert the problem to an initial value representation, and then run ensembles of trajectories~\cite{Heller81, Herman84, Heller91b, Kay94, Miller01}.  Recently,  along this vein there have been some ideas posited for dealing with multidimensional systems, such as running initial conditions just along the phase space direction leading to the most unstable dynamics~\cite{Kocia14b,Kocia15}  and a ``divide and conquer'' scheme~\cite{Ceotto17}.  However, there is a more comprehensive version of the most unstable direction idea within a general framework, not initially motivated by wave packet or coherent state propagation~\cite{Sala16}.  This approach, called the anisotropic method, was introduced for classically chaotic systems with multiple positive Lyapunov exponents.   It can be modified and adapted for our purposes.  

In this paper, the focus is on developing techniques for solving the boundary value problem for multi-dimensional Gaussian wave packets. The technique is applied in the context of propagating a bosonic many-body coherent state, although it could have just as easily been applied to a multidimensional Hamiltonian of the type used to describe molecular dynamics or other systems.  It gives an account of the calculation methods used in~\cite{Tomsovic18b,Schlagheck18,Ullmo18}.  For wave packets, the Lagrangian manifolds necessarily contain complex momenta and positions and thus the intersections are complex saddle points~\cite{Huber88}, whose properties are determined by their respective saddle trajectories.  Thus, the full phase space of a system with $N$ degrees of freedom is $2N$ complex dimensions ($4N$ parameters to define a phase point).  The Lagrangian manifold of a Gaussian wave packet is known to be an $N$-dimensional complex hyperplane~\cite{Huber88}.  The boundary value problem to be solved first requires the determination of the remaining $2N$ free parameters that define each solution point on the manifolds satisfying the boundary conditions.  Then, only the saddles with physical relevance are to be kept, and the rest must be thrown away.  

Even though, it is more or less hopeless to perform a full many-dimensional `blind' search for solutions, it can  be extremely helpful to account for the asymptotic structure in the flow of Hamiltonian dynamical systems.  This enables one to orient the search in such a way as to rely on a much lower dimensional subspace of the full system as in the spirit of~\cite{Kocia14b,Kocia15,Sala16}.  The approach derived below is similar to the method of decomposing the tangent space used in numerical calculations of Lyapunov exponents~\cite{Gaspard98,Ott02}.  It relies on the stability matrix of a wave packet's central trajectory, its transpose, and the shape parameters of the wave packet.  The result gives a spectrum (eigenvalues) of expansion and contraction rates and their directions (eigenvectors) in the space of initial conditions.  The method works regardless of the nature of the dynamics, be they integrable, chaotic, or some mixture.

For Hamiltonian dynamical systems, this process immediately identifies half of the spatial dimensions as being irrelevant due to their contractional behavior; i.e.~trajectories whose initial conditions are aligned along these directions approach each other as they propagate and and cannot be responsible for different sets of saddles or provide any new information.  The remaining eigenvectors form a reduced dimensional initial condition search space containing all the saddles.  For large classes of dynamical systems, it is possible to continue this analysis and further reduce the search space dimensionality.

For example, order the eigenvectors by their rate of expansion (according to their associated eigenvalues) proceeding from greatest to least.  For systems with many degrees of freedom, a significant gap in the expansion rates between the most unstable eigenvectors and all the rest may appear.  If so, it turns out to be possible to locate the relevant saddle trajectories to intermediate time scales using initial conditions aligned along just the most rapidly expanding directions.  Symmetries can play an important role, leading to rather interesting behaviors, that also lead to further reductions in the search space necessary to find all the relevant saddles.  

Furthermore, it seems to happen that even highly unstable directions do not necessarily lead to the creation of additional saddles.  Apparently, the trajectories can separate rapidly, yet not lead to additional transport pathways on the time scales in which one can follow the stability of the trajectories accurately.  Apparently each unstable direction can be checked individually to determine whether to retain it in the search space, and thus, combined with the previous condition, the minimum dimensional search space accounts only for those eigenvectors corresponding to the greatest instability (above any gap that may exist) and those that create additional pathways.

The further complicating factor of determining whether a solution to this boundary value problem should be kept can be connected with the ability of complex classical mechanics to create the possibility of `runaway' trajectories, i.e.~trajectories that attain infinite momentum in finite time~\cite{Huber88}.  They are responsible for branch cuts in action functions and it is necessary to throw away solutions on the wrong side of the cuts, a Stokes phenomenon.  Progress has been made on this issue and we follow the technique described in~\cite{Pal16}.  That work showed that the contributing saddle points could be put into a one-to-one correspondence with individual bundles of similarly behaving real trajectories that represent unique transport pathways.  Choosing one representative trajectory from each bundle as a seed trajectory coupled with a Newton-Raphson method quickly converged to the set of contributing complex saddle trajectories.  

This has three very desirable features.  First, it allows one to work in a $2N$-dimensional real phase space to search for transport pathways instead of with an $N$-dimensional complex manifold embedded in a $2N$-dimensional complex phase space.  Second, it provides a criterion for how to cut off the phase space volume necessary to search.  The saddles associated with the region exterior to this domain contribute as negligibly as one chooses, depending on the volume cut off.  This reduces the problem to one in which  intuition and knowledge of real classical dynamics is sufficient eventually to solve the complex boundary value problem.  Finally, all the saddles found this way contribute, thus entirely avoiding the necessity of a scheme or criterion to determine whether a saddle must be kept and the problem of wasting effort on saddles that must be thrown away. 

This paper is structured as follows, the next section introduces the Gaussian wave packets, their Lagrangian manifolds, their Wigner transforms, and summarizes the Newton-Raphson scheme that allows one to focus on real classical transport.  The following section discusses the theory behind reducing the dimensionality of the search to a manageable level.  After that a Bose-Hubbard model system with several degrees of freedom is introduced and saddle trajectories are identified.  The role of symmetries is discussed.  In an addition section, it is shown how the above mentioned spectrum and associated eigenvectors reflect the various dynamical regimes in different parts of the available phase space.  The summary and conclusions consider the  strengths and difficulties associated with the method.

\section{Gaussian wave packets/coherent states: saddle point conditions}
\label{gwp}

As mentioned briefly in the introduction, multidimensional Gaussian wave packets show up in many subfields of physics and have become extremely important tools for understanding a wide range of phenomena.  In addition, the projection into configuration space of a coherent state describing a bosonic many-body system of the form
\begin{equation}
\label{cs}
| z \rangle = \exp \left(-\frac{\left| z \right|^2}{2} + z \hat a^\dagger \right)| 0\rangle 
\end{equation}
results in a Gaussian wave packet~\cite{Glauber63}, and the parameters of the coherent state are straightforwardly mapped onto those of the wave packet; see Appendix~\ref{cswp}.  As Gaussian wave packets are extremely important in and of themselves, and it is possible to create a more general wave packet than the one that follows from this particular coherent state form, the development of the theory ahead is given in terms of the most general wave packet.   If needed, translating all of the results back into the language of coherent states is possible in a straightforward way, i.e.~$z$ can be mapped onto momentum and position centroids, and the ground state determines the shape parameters.

\subsection{Gaussian wave packets}
\label{gwp1}

A Gaussian wave packet has a number of parameters needed in order to specify it uniquely; we label the entire set with a Greek letter, such as $\alpha$ or $\beta$.  Thus, the real mean momenta and positions are labelled $(\vec p_\alpha, \vec q_\alpha)$, and the matrix ${\bf b}_\alpha$ describes all the possible shape parameters.  It must be a symmetric matrix diagonalizable by an orthogonal matrix with eigenvalues whose real parts are positive in order to be square integrable.  If ${\bf b}_\alpha$ is complex, then the wave packet is sometimes called a ``chirped'' wave packet, i.e.~one in which the speed of phase oscillations linearly increases or decreases across its width.  We choose the phase convention and $\hbar$-dependence such that 
\begin{eqnarray}
\label{wavepacket}
\phi_\alpha(\vec x) &=& \exp\left[ - \left(\vec x - \vec q_\alpha \right) \cdot \frac{{\bf b}_\alpha}{2\hbar} \cdot \left(\vec x - \vec q_\alpha \right) +\frac{i}{\hbar} \vec p_\alpha \cdot \left(\vec x - \vec q_\alpha \right)\right] \nonumber \\ && \times \left[\frac{{\rm Det}\left({\bf b}_\alpha+{\bf b}^*_\alpha\right)}{(2\pi\hbar)^N}\right]^{1/4}
\end{eqnarray}
which represents a different phase convention than that implied by Eq.~(\ref{cs}), but that is accounted for properly when applied to the Bose-Hubbard model ahead. Implicitly the right vectors are column vectors and the left vectors are row vectors.  The $\hbar$ scaling chosen ensures that $\hbar$ determines the volume occupied by the wave packet, and its overall shape is completely independent of $\hbar$.  The dual of this wave packet follows by the complex conjugation of ${\bf b}_\alpha$ and the sign change in front of the momentum term.  The notation for an evolving wave packet follows as $\phi_\alpha(\vec x;t)$, but in general, it ceases to maintain a Gaussian form for $t>0$. 

Assume the existence of a classical Hamiltonian, which can be analytically continued to complex phase space variables $H=H(\vec {\cal p},\vec {\cal q};t)$, and a well defined corresponding quantum Hamiltonian, $\hat H= \hat H(\frac{\hbar}{i}\partial /\partial \vec x, \vec x;t)$.  They govern the classical and quantum dynamics, respectively.  Two very basic dynamical quantities of interest are given by the evolving wave packet itself, $\phi_\alpha(\vec x;t)$, and so-called correlation functions
\begin{eqnarray}
\label{ac}
{\cal A}_{\beta\alpha}(t) &=& \int_{-\infty}^\infty {\rm d}\vec x\ \phi^*_\beta(\vec x) \phi_\alpha(\vec x;t) \nonumber \\
{\cal C}_{\beta\alpha}(t) &=& \left| {\cal A}_{\beta\alpha}(t)\right|^2 
\end{eqnarray}
where, if the set of parameters labelled by $\beta$ and $\alpha$ are equal, then ${\cal C}_{\alpha\alpha}(t)$ is called the autocorrelation function.  A matrix element of the Feynman path integral in a coherent state representation would be equivalent to the amplitude, ${\cal A}_{\beta\alpha}(t)$, of a correlation function.

\subsection{Lagrangian manifolds}

The Lagrangian manifold for a wave packet is the set of all complex positions and conjugate momenta  $(\vec {\cal p}, \vec {\cal q})$ satisfying the equations~\cite{Huber88}
\begin{equation}
\label{constraints}
 {\bf b}_\alpha \cdot \left( \vec {\cal q} - \vec q_\alpha\right) + i \left( \vec {\cal p} - \vec p_\alpha\right) = 0
\end{equation}
Notice that the manifold has no dependence on $\hbar$.  This gives an $\hbar$ independent boundary value problem to solve, which explains the placement choice of $\hbar$ in  Eq.~(\ref{wavepacket}).    A dual wave packet with a possibly different parameter set leads to the modified Lagrangian manifold equations
\begin{equation}
\label{constraintsbra}
 {\bf b}^*_\beta \cdot \left( \vec {\cal q} - \vec q_\beta\right) - i \left( \vec {\cal p} - \vec p_\beta\right) = 0
\end{equation}

The semiclassical approximation~\cite{Maslov81} relies on saddle points whose properties are given by trajectories with initial conditions, $(\vec {\cal p}_0, \vec {\cal q}_0)$, that lie on the initial manifold and after propagation of a time $t$, $(\vec {\cal p}_t, \vec {\cal q}_t)$, end up on the final manifold.  Thus for correlation functions, the boundary value problem is to find all contributing solutions of the equations
\begin{eqnarray}
\label{sadcond}
{\bf b}_\alpha \cdot \left( \vec {\cal q}_0 - \vec q_\alpha\right) + i \left( \vec {\cal p}_0 - \vec p_\alpha\right) &=& 0 \nonumber \\
{\bf b}^*_\beta \cdot \left( \vec {\cal q}_t - \vec q_\beta\right) - i \left( \vec {\cal p}_t - \vec p_\beta\right) &=& 0
\end{eqnarray}
as a function of $t$.  If interest is in the evolving wave packet in the configuration space representation, then the final Lagrangian manifold must be the one associated with $\langle \vec x|$, and the second set of equations is replaced by
\begin{equation}
\vec {\cal q}_t = \vec x
\end{equation}
where $\vec {\cal p}_t$ can be anything.  We will call the trajectories satisfying these conditions saddle trajectories.

Generally speaking, excluding harmonic oscillators (or rather systems with linear Hamilton's equations), there appear to be an infinity of solutions to these equations, almost all of which either must be excluded for reasons mentioned in the introduction, or are irrelevant because they contribute so little that they are vastly smaller than the errors involved in making a semiclassical approximation.  The goal then is to find all the saddle trajectories that must be included and contribute sufficiently.  The number of relevant saddles grows at least linearly with increasing time for integrable dynamical systems and exponentially for chaotic ones.  If for no other reason, this gives a practical upper limit to the length of propagation time conceivable with semiclassical methods.  The domain around each saddle point for which the Newton-Raphson scheme can work shrinks accordingly.  Eventually, the search has to be carried out on too fine a scale to be practical.

Interestingly, for wave packets any initial condition $(\vec {\cal p}_0, \vec {\cal q}_0)$ on the Lagrangian manifold can play the role of the real centroid $(\vec p_\alpha, \vec q_\alpha)$ in Eq.~(\ref{wavepacket}), i.e.~the interchange leaves the spatial dependence of the wave packet invariant.  However, the normalization constant has to be redefined to
\begin{eqnarray}
\label{norm1}
{\cal N}_\alpha^0 &=& \left[\frac{{\rm Det}\left({\bf b}_\alpha+{\bf b}^*_\alpha\right)}{(2\pi\hbar)^N}\right]^{1/4} \exp \left[ \frac{i}{\hbar}\left( \vec{\cal p}_0 \cdot \vec{\cal q}_0 - \vec p_\alpha \cdot \vec q_\alpha\right) + \right.  \nonumber \\
&& \left.  \vec {\cal q}_0 \cdot \frac{{\bf b}_\alpha}{2\hbar} \cdot \vec {\cal q}_0 - \vec q_\alpha \cdot \frac{{\bf b}_\alpha}{2\hbar} \cdot \vec q_\alpha \right. \Big]
\end{eqnarray}
in order to preserve the normalization and phase convention.  The similar substitution for correlation functions of the trajectory endpoint is given by
\begin{eqnarray}
\label{norm2}
{\cal N}_\beta^t &=& \left[\frac{{\rm Det}\left({\bf b}_\beta+{\bf b}^*_\beta\right)}{(2\pi\hbar)^N}\right]^{1/4} \exp \left[ -\frac{i}{\hbar}\left( \vec{\cal p}_t \cdot \vec{\cal q}_t - \vec p_\beta \cdot \vec q_\beta\right) + \right.  \nonumber \\
&& \left.  \vec {\cal q}_t \cdot \frac{{\bf b}^*_\beta}{2\hbar} \cdot \vec {\cal q}_t - \vec q_\beta \cdot \frac{{\bf b}^*_\beta}{2\hbar} \cdot \vec q_\beta \right. \Big]
\end{eqnarray}
This substitution and modified normalization constants can be used to simplify the final form of the semiclassical (saddle point) approximation.

\subsection{Real classical transport and saddle trajectories}

It was shown in~\cite{Pal16} that there is a one-to-one correspondence between real classical transport pathways (bundles of like-behaving trajectories) and the relevant complex saddle trajectories.  It suffices to start with a seed trajectory given by a single representative trajectory for a specific pathway and use a Newton-Raphson scheme to locate the corresponding and contributing saddle trajectory.    This scheme has the three highly desirable main consequences mentioned in the introduction.

Here, we give for completeness the equations that arise in the Newton-Raphson scheme~\cite{Pal16}.  Considering the phase space in the neighborhood of a seed trajectory, it is useful to define $\delta \vec {\cal p}_t =  \vec {\cal p} - \vec {\cal p}_t$ and $\delta \vec {\cal q}_t =  \vec {\cal q} - \vec {\cal q}_t$.  The stability matrix ${\bf M}_t$ describes how neighboring trajectories shift relative to this seed trajectory.  Thus, 
\begin{equation}
\left( \begin{array}{c} \delta \vec {\cal p}_t \\ \delta \vec {\cal q}_t \end{array} \right) =  \left( \begin{array}{c} \bf{M_{11}} \\ \bf{M_{21}} \end{array} \begin{array}{c} \bf{M_{12}} \\ \bf{M_{22}} \end{array} \right)
\left( \begin{array}{c} \delta \vec {\cal p}_0 \\ \delta \vec {\cal q}_0 \end{array} \right) 
\label{delta}
\end{equation}
The seed orbit most likely does not satisfy the boundary value problem and in the case of correlation functions instead gives
\begin{eqnarray}
{\bf b}_\alpha \cdot \left( \vec {\cal q}_0 - \vec q_\alpha\right) + i \left( \vec {\cal p}_0 - \vec p_\alpha\right) &=& \vec {\cal c}_0 \nonumber \\
{\bf b}^*_\beta \cdot \left( \vec {\cal q}_t - \vec q_\beta\right) - i \left( \vec {\cal p}_t - \vec p_\beta\right) &=& \vec {\cal c}_t
\end{eqnarray}
Combining these and the stability equations, it is possible to solve for the change in initial conditions needed to approach the saddle trajectory.  This gives
\begin{equation}
\label{nr}
\begin{array}{l}
 \vec{\cal{p}_0}^\prime = \vec{\cal{p}}_0 + i{\bf b}_\alpha \cdot {\cal D} \cdot \left[ \left( {\bf b}^*_\beta \cdot{\bf M_{22}} - i  {\bf M_{12}} \right) \cdot {\bf b}_\alpha^{-1} \cdot \vec  {\cal c}_0 - \vec  {\cal c}_t \right] \\
 \vec{\cal{q}_0}^\prime = \vec{\cal{q}}_0 - {\cal D} \cdot \left[ \left( {\bf M_{11}}+ i {\bf b}^*_\beta \cdot {\bf M_{21}} \right) \cdot \vec {\cal c}_0 + \vec {\cal c}_t \right]
\end{array}
\end{equation}
where
\begin{equation}
{\cal D}^{-1} =  {\bf M_{11}}\cdot {\bf b}_\alpha + {\bf b}^*_\beta \cdot {\bf M_{22}} + i {\bf b}^*_\beta\cdot  {\bf M_{21}}\cdot {\bf b}_\alpha - i{\bf M_{12}} 
\end{equation}
If the interest is in calculating the propagating wave packet itself, as opposed to some correlation function, the equations are slightly simplified to give
\begin{equation}
\begin{array}{l}
 \vec{\cal{p}_0}^\prime = \vec{\cal{p}}_0 + i{\bf b}_\alpha \cdot {\cal D} \cdot \left[ {\bf M_{22}}  \cdot {\bf b}_\alpha^{-1} \cdot \vec {\cal c}_0 - \vec {\cal c}_t \right] \\
 \vec{\cal{q}_0}^\prime = \vec{\cal{q}}_0 -{\cal D} \cdot \left[ i {\bf M_{21}}  \cdot \vec {\cal c}_0 + \vec {\cal c}_t \right]
\end{array}
\end{equation}
with
\begin{equation}
\label{det2}
{\cal D}^{-1} =  {\bf M_{22}} + i {\bf M_{21}}\cdot {\bf b}_\alpha  \qquad \vec {\cal q}_t - \vec x = \vec {\cal c}_t
\end{equation}
These equations are used iteratively to converge to a contributing saddle point.  It suffices to find a single point within the domain of convergence for each saddle, which is what the seed trajectories provide.

\subsection{Saddle families}
\label{family}

In a continuous time dynamical system, i.e.~as opposed to dynamical mappings, each saddle gives rise to a one parameter family of saddles.  As $t$ changes continuously, the saddle trajectory's initial conditions change continously as well.  Barring orbit bifurcations and crossing Stokes surfaces (which becomes exceedingly unlikely in the $\hbar\rightarrow 0$ limit), it is possible to predict how the initial conditions change using Eq.~(\ref{sadcond}) and Hamilton's equations.  Consider a saddle trajectory that contributes at exactly time $t$, thus satisfying Eq.~(\ref{sadcond}).  It's initial condition lies on the initial Lagrangian manifold, and its propated endpoint on the final one.  If however, the propagation time is slightly (differentially) altered, the endpoint is no longer on the final manifold.  Using Hamilton's equations for a time shift $\delta t$, the altered endpoint is located at, 
\begin{eqnarray}
\vec {\cal q}_{t+\delta t} &=& \vec {\cal q}_t + \frac{\partial H}{\partial \vec {\cal p}_t} \delta t \nonumber \\
\vec {\cal p}_{t+\delta t} &=& \vec {\cal p}_t - \frac{\partial H}{\partial \vec {\cal q}_t} \delta t
\end{eqnarray}
The Newton-Raphson scheme of the previous section can be applied to find the shift in initial conditions that would restore the saddle point conditions for the new time.  Since the initial point begins on the initial manifold, $\vec c_0=0$, but the shift of the final point means that $\vec c_t\ne 0$.  Following the same kind of algebra leading to Eq.~(\ref{nr}) gives the initial condition expressions for the saddle trajectory, which contributes at $t+\delta t$, 
\begin{equation}
\begin{array}{l}
\label{tshift}
 \vec{\cal{p}}_0^{\{t+\delta t\}} = \vec{\cal{p}}_0^{\{t\}} - i{\bf b}_\alpha \cdot {\cal D} \cdot {\bf b}^*_\alpha \cdot \left( \frac{\partial H}{\partial \vec {\cal p}_t} +i \frac{\partial H}{\partial \vec {\cal q}_t} \right) \delta t \\
 \vec{\cal{q}}_0^{\{t+\delta t\}} = \vec{\cal{q}}_0^{\{t\}} - {\cal D} \cdot {\bf b}^*_\alpha \cdot \left( \frac{\partial H}{\partial \vec {\cal p}_t} +i \frac{\partial H}{\partial \vec {\cal q}_t} \right) \delta t
\end{array}
\end{equation}
A similar expression results for the case in which the quantity of interest is the propagating wave function with the matrix ${\bf b}^*_\alpha$ replaced by unity and the simpler determinant $\cal D$ of Eq.~(\ref{det2}).  The structure of these equations involving the gradient of the Hamiltonian is linked to the fact that the direction of initial condition variation is along the maximal change of  (perpendicular to) the energy surface in an autonomous dynamical system.

\begin{figure}[tbh]
\includegraphics[width=8.5 cm]{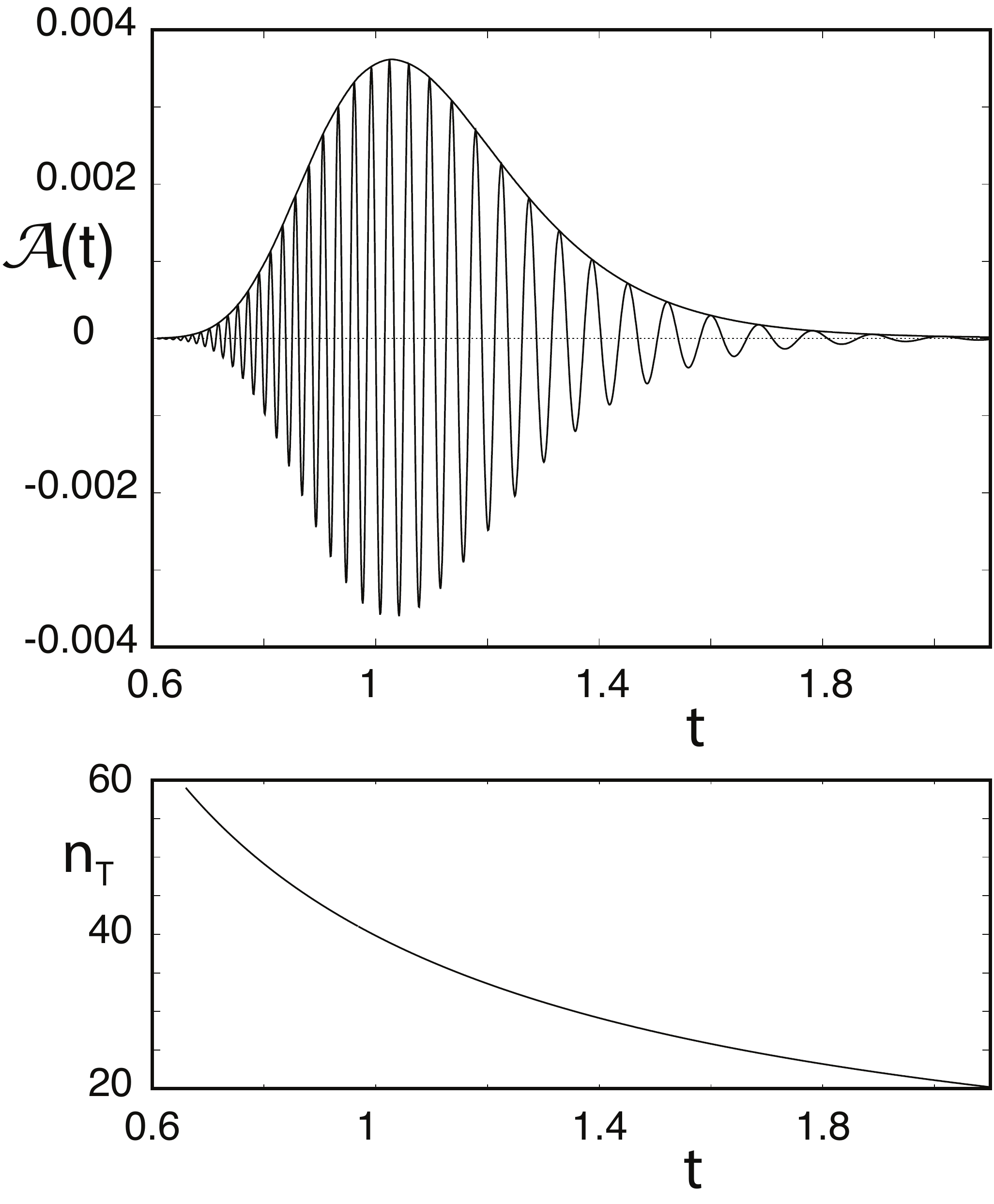} 
\caption{Typical saddle family characteristics.  The oscillating curve in the upper panel is the real part of ${\cal A}(t)$ for one particular saddle family, and the envelope is the absolute value.  There is a faster phase oscillation at short times decreasing as time increases corresponding to changes in the complex saddle trajectory with time.  Each saddle family member has an energy and total particle number ($n_T$) surface to which it belongs.  At short times, the real parts of the energy and particle number of the saddle trajectory are greater than the energy and particle number expectation values of the wave packet, and at longer times they are less than the expectation values.  The saddle family's peak contribution occurs near where the real parts of the energy and $n_T$ equals the energy and total particle number (here $<n_T>=40$) expectation values of the wave packet.  This saddle family is taken from an example of the Bose-Hubbard model defined ahead in Sect.~\ref{bhms}.  \label{fig1}} 
\end{figure}
The modified initial conditions of Eq.~(\ref{tshift}) can be used as a seed for the Newton-Raphson scheme of the previous section to construct the entire saddle trajectory family that forms a continuous time contribution to the evolving wave packet or correlation function.  An example from the Bose-Hubbard model introduced in Sect.~\ref{bhms} is shown for illustration purposes in Fig.\ref{fig1}.  Generally speaking, there is a peak contribution time for a saddle family corresponding to a saddle trajectory possessing an energy and particle number close to the mean of the initial wave packet.  Earlier and later in time, the saddle trajectory moves further away from this energy and particle number surface and the contribution decays, thus creating a time window in which it contributes significantly.  It suffices to search for a single real transport pathway seed on the energy and particle number surface of the trajectory defined by $(\vec p_\alpha, \vec q_\alpha)$, locate a saddle, and from there obtain the contribution of the entire family through repeated use of Eq.~(\ref{tshift}).  In practice, the convergence appears to be superior (computationally faster and fewer convergence problems) if constructing the entire saddle family this way than to find real seed trajectories as a continuous function of time.

\section{Identifying real classical transport pathways}
\label{transport}

\subsection{Wigner transform}
\label{wt}

The key for identifying classical transport pathways is to start with the phase space image of a wave packet under the Wigner transform.  This gives a multidimensional Gaussian density of phase points in a classical phase space to consider.  This image is given by
\begin{equation*}
{\cal W}(\vec p, \vec q) = \frac{1}{(2\pi\hbar)^{N}} \int_{-\infty}^\infty {\rm d} \vec x \ {\rm e}^{i \vec p \cdot \vec x/\hbar} \phi_\alpha \left(q-\frac{\vec x}{2}\right)  \phi^*_\alpha \left(q+\frac{\vec x}{2}\right) \end{equation*}
\begin{equation}
= \left(\pi \hbar \right)^{-N} \exp \left[ - \left(\vec p - \vec p_\alpha, \vec q - \vec q_\alpha \right) \cdot \frac{{\bf A}_\alpha}{\hbar} \cdot \left(\vec p - \vec p_\alpha, \vec q - \vec q_\alpha \right) \right] 
\end{equation}
where ${\bf A}_\alpha$ is
\begin{equation}
\label{mvg}
{\bf A}_\alpha = \left(\begin{array}{cc}
{\bf  c^{-1}} & {\bf  c}^{-1} \cdot {\bf  d}  \\
 {\bf  d} \cdot {\bf c}^{-1} & {\bf c} + {\bf  d} \cdot {\bf c}^{-1} \cdot {\bf  d} \end{array}  
\right)   \qquad {\rm Det}\left[ {\bf A}_\alpha \right] =1
\end{equation}
with the association 
\begin{equation}
\label{mvgwf}
{\bf b}_\alpha = {\bf c} + i {\bf d}
\end{equation}
The $ 2N \times 2N$ dimensional matrix ${\bf A}_\alpha$ is real and symmetric.  If $\bf b_\alpha$ is real, there are no covariances between $\vec p$ and $\vec q$; i.e.~the wave packet is not chirped.  The off-diagonal blocks of the matrix ${\bf A}_\alpha$ disappear.  

Ahead it is very useful to know that ${\bf A}_\alpha$ can be inverted analytically.  The inverse is given by~\cite{Lu02}
\begin{equation}
{\bf A}_\alpha^{-1} = \left(\begin{array}{cc}
{\bf c} + {\bf  d} \cdot {\bf c}^{-1} \cdot {\bf  d} & - {\bf  d} \cdot {\bf c}^{-1}  \\
- {\bf  c}^{-1} \cdot {\bf  d} & {\bf  c^{-1}} \end{array}  \right) 
\end{equation}
Since it is necessary to calculate $\bf c^{-1}$ to determine ${\bf A}_\alpha$, its inverse is determined with no further effort.

\subsection{Local evolution of Gaussian densities}
\label{egd}

Consider any constant density contour of the Wigner transform of the initial wave packet as a set of initial conditions.  It must have some kind of hyper-elliptical shape described by the equation
\begin{equation}
\label{ellipse0}
r^2 = \left( \delta \vec p_0 , \delta \vec q_0 \right) \cdot  \frac{{\bf A}_\alpha}{\hbar} \cdot \left(\delta \vec p_0 , \delta \vec q_0 \right)
\end{equation}
where $(\delta \vec p_0, \delta \vec q_0) = (\vec p_0 - \vec p_\alpha, \vec q_0 - \vec q_\alpha)$, where $(\vec p_0, \vec q_0)$ belong to a set of points on the hyper-elliptical surface satisfying the equation.  Locally, within a linearizable regime (small enough $r$), the dynamics to time $t$ distorts the hyper-ellipse to a new one
\begin{equation}
\label{ellipse}
r^2 = \left( \delta \vec p_t , \delta \vec q_t \right) \cdot  \frac{{\bf A}_\alpha (t)}{\hbar} \cdot \left(\delta \vec p_t , \delta \vec q_t \right)
\end{equation}
Recalling the information given by the stability matrix of the central trajectory $(\vec p_0, \vec q_0)=(\vec p_\alpha, \vec q_\alpha)$ identifies the evolution of ${\bf A}_\alpha$ with $t$.  Inserting unity of the form $\mathbb{1}=  {\bf M}_t^{-1}  {\bf M}_t$ and its transpose appropriately into Eq.~(\ref{ellipse0}) gives
\begin{eqnarray}
r^2 &=& \left( \delta \vec p_0 , \delta \vec q_0 \right)  \cdot {\bf M}_t^T  \cdot{ {\bf M}_t^{-1}}^T \cdot \frac{{\bf A}_\alpha}{\hbar} \cdot {\bf M}_t^{-1} \cdot {\bf M}_t \cdot \left(\delta \vec p_0, \delta \vec q_0 \right) \nonumber \\
&=& \left( \delta \vec p_t , \delta \vec q_t \right) \cdot {{\bf M}_t^{-1}}^T \cdot \frac{{\bf A}_\alpha}{\hbar} \cdot {\bf M}_t^{-1}  \cdot \left(\delta \vec p_t, \delta \vec q_t \right)
\end{eqnarray}
and, thus, necessarily one has the identification
\begin{equation}
{\bf A}_\alpha (t) = {{\bf M}_t^{-1}}^T \cdot {\bf A}_\alpha \cdot {\bf M}_t^{-1}
\end{equation}
This is a real symmetric matrix (also with unit determinant) which can be diagonalized by an orthogonal transformation.  Its eigenvalues and eigenvectors contain all the information necessary to enable a targeted search for saddle trajectories.  For convenience, we work with the inverse, which has the exact same set of eigenvectors, i.e.
\begin{equation}
\Lambda = {\cal O} {\bf A}_\alpha^{-1}(t) {\cal O}^{-1} = {\cal O} {\bf M}_t \cdot {\bf A}_\alpha^{-1} \cdot {\bf M}_t^T {\cal O}^{-1}
\end{equation}
and the set of inverse eigenvalues, $\{\lambda_{j,\pm}\}$.  The determinant of ${\bf A}_\alpha^{-1}(t)$ is unity and the eigenvalues come in pairs here labelled by $j=1,...N$, one expanding, $\lambda_{j,+} > 0$, one contracting, $\lambda_{j,-} <0$ ($\lambda_{j,+}=\lambda_{j,-}^{-1}$).  Similar constructs have been used in the calculation of the various Lyapunov exponents of a multidimensional chaotic dynamical system, where the process is discussed as a decomposition of the tangent space~\cite{Gaspard98,Ott02}.

\subsection{Asymptotic structure}
\label{as}

For a large class of systems and initial states, it will turn out that most of the degrees of freedom do not need to be part of the search.  Dynamical systems have a great deal of structural organization in their phase spaces that is revealed asymptotically in time by the ${\bf A}_\alpha^{-1}(t)$ matrix.  Denote the eigenvectors corresponding to the set of $\lambda_{j,+}$ as $(\delta \vec p_t, \delta \vec q_t)_j$.  Each eigenvector signifies the final direction of a set of initial conditions along a line, which separated at the rate $\lambda_{j,+}$.  One wishes to know which set of initial conditions in the neighborhood of $(\vec p_\alpha, \vec q_\alpha)$ ends up evolving into the eigenvector $(\delta \vec p_t, \delta \vec q_t)_j$.  Using the definition of the stability matrix, it turns out to be the direction of initial conditions given by
\begin{equation}
\left( \begin{array}{c} \delta \vec p_0 \\ \delta \vec q_0 \end{array} \right)_j =  {\bf M}_t^{-1}
\left( \begin{array}{c} \delta \vec p_t \\ \delta \vec q_t \end{array} \right)_j
\label{delta2}
\end{equation}
The vector of initial conditions depends on the length of propagation time used to generate the ${\bf A}_\alpha^{-1}(t)$ matrix.  However, for large times, each vector of initial conditions converges to a stable direction, and becomes essentially independent of time.  If a propagation time too short is selected, then the initial condition vectors will not have stabilized, i.e.~converged to the directions of interest.  On the other hand, propagation that covers too long a time period risks losing accuracy and will eventually cause numerical problems.  Here, we construct ${\bf A}_\alpha^{-1}(\tau)$, $t=\tau$, for a long intermediate time scale within the appropriate time range and use its eigenvectors to determine the most important degrees of freedom to sample.  This is done once at the very beginning to initiate the process of finding real seed trajectories as a function of time.  An indication of how to arrive at a reasonable time scale $\tau$ is given in the next subsection.

Only the $N$ eigenvectors associated with the eigenvalues greater than unity need to be considered.  Trajectories linked by a contracting direction only approach each other, and evolve similarly.  If a trajectory belongs to a bundle corresponding to a classical pathway, so will all of its neighbors along the $N$ contracting degrees of freedom, i.e.~the $N$-dimensional manifold described by the $N$ contracting eigenvectors.  

\subsection{Distinguishing shearing and exponential stretching, and a reasonable value of $\tau$}
\label{distinguish}

It is not necessary to search in the direction that maximizes the change of energy.  This is related to the saddle families discussed in Sect.~\ref{family} and this direction is already accounted for by the technique described in that section.  Thus, a targeted search for seed trajectories can be immediately reduced to an $N-1$ dimensional parameter search of initial conditions in a real phase space without any loss of generality (assuming the omission of contracting directions).  The associated eigenvector needs to be identified in order to avoid sampling in that direction.  As it must be associated with a shearing in the dynamics, it cannot be associated with the exponential stretching of instability.

There is a simple trick that often suffices to identify this eigenvector quickly, and which helps identify whether one has reached a sufficiently asymptotic propagation time, $\tau$ (this does not work for a harmonic oscillator where there is no shearing in the dynamics).  The logic follows by considering free particle motion in a single degree of freedom.  Let ${\bf A}_\alpha$ and the mass be unity and irrelevant for this purpose.  The stability matrix times its transpose is
\begin{equation}
{\bf M}_\tau \cdot {\bf M}_\tau^T = \left(  
\begin{array}{cc}
1 & 0 \\
\tau & 1
\end{array}
\right) \cdot 
\left(  
\begin{array}{cc}
1 & \tau \\
0 & 1
\end{array}
\right) = \left(  
\begin{array}{cc}
1 & \tau \\
\tau & 1 + \tau^2
\end{array}
\right) 
\end{equation}
with large eigenvalue 
\begin{equation}
\lambda_+(\tau) = 1 + \frac{\tau^2}{2} +\frac{1}{2}\sqrt{\tau^4+4\tau^2} \approx \tau^2
\end{equation}
where the approximate result applies only if $\tau$ is large enough.  In a multidimensional system with more complicated dynamics, quadratic dependence of this eigenvalue is an indicator that the asymptotic structure of its Hamiltonian flow has emerged.  Therefore, if one calculates the spectrum, $\{\lambda_{j,\pm}\}$, for time $\tau$ and $2\tau$ sufficiently large, there must be an eigenvalue for which $\lambda_{j,+} (2\tau) = 4\lambda_{j,+}(\tau)$.  If there is only one, then its eigenvector must be perpendicular to the energy surface.  If there are none, then one has not reached the asymptotic regime desired and $\tau$ must be increased.  If there are multiple eigenvalues respecting this relation, then one can calculate the energy along the associated multiple eigenvectors to determine which maximally shifts the energy or calculate the gradient of the Hamiltonian at the wave packet centroid and compare to the relevant eigenvectors.

Unstable degrees of freedom behave very differently.  As their eigenvalues behave exponentially in time, one expects fully unstable directions to satisfy, $\lambda_{j,+} (2\tau) = \lambda_{j,+}^2(\tau)$.  In practice, one finds a factor of unity (no stretching at all) or square relations as limiting possibilities, and the various eigenvalue behaviors lie in between these cases.  In fact, in the calculations performed ahead, only one eigenvalue followed the factor four relation and it was unnecessary to calculate the gradient of the energy surface and compare it to an eigenvector.

\subsection{Determining the initial condition sampling space}

Of the remaining $N-1$ dimensional phase space of initial conditions of relevance to searching for classical transport pathways, consider the the largest eigenvalue first, denote it $\lambda_{1,+}$ and its associated vector of initial conditions $(\delta \vec p_0, \delta \vec q_0)_1$.   It gives a very particular coordinate direction of initial conditions in which to search for the earliest appearing real transport pathways.  It should be emphasized that the range of initial conditions along this vector are chosen to fully span the breadth of the initial wave packet's Wigner transform Gaussian density, i.e.~as many standard deviations as desired.  They are not limited to the linearizable regime used to identify this direction.  The line of initial conditions is propagated long enough in time to become highly stretched, nonlinear, and repeatedly folded into an extremely complicated shape, i.e.~it is used far beyond the linearizable regime that was used to identify the direction. 

If the second largest eigenvalue $\lambda_{2,+}$ is not too much smaller than $\lambda_{1,+}$, then it is likely necessary to add another search direction for saddle trajectories, i.e.~the phase space plane of initial conditions defined by the first and second vectors $(\delta \vec p_0, \delta \vec q_0)_1$ and $(\delta \vec p_0, \delta \vec q_0)_2$.  One could continue in this way to successively higher dimensions until the most relevant initial conditions are included in the search.  However, it appears that sometimes an unstable direction does not generate additional saddles for the dynamical quantity of interest.  For example, concerning the autocorrelation function, this would mean that even though the various initial conditions lead to rapidly separating trajectories, away from the central trajectory along this direction they do not result in additional returning trajectories within the time frame of interest.  In fact, for the Bose-Hubbard model of Sect.~\ref{bhms}, in some cases even when combined with another part of the subspace, which does generate transport pathways leading to saddles, new saddles seem not to appear.   This could be true for other dynamical systems as well.  Therefore, one can check each expanding direction individually as an indicator of which collection of eigenvectors (subspace of initial conditions) is absolutely necessary for an exhaustive saddle search, and one can use this as a starting point for a minimal search subspace.  However, we are not currently aware of any guarantee that this is always going to turn out to be sufficient.

We recognize that in practice it may not really be all that practical to continue beyond say, $3$ dimensions.  Nevertheless, for a broad class of dynamical systems and wave packets, even possessing many degrees of freedom, this is sufficient for the purpose of constructing the semiclassical prediction for correlation functions.  Some examples with up to $8$ degrees of freedom are shown in Sect.~\ref{bhms}.

\subsection{Finding seed trajectories}
\label{seed}

With the sampling space determined, the goal is reduced to identifying a single seed trajectory for each unique pathway.  One simple idea is to define a function of the initial conditions in the sampling space for which one can search for local minima.  Consider the correlation function as a concrete example.  The Wigner transform of the final state has a centroid $(\vec p_\beta, \vec q_\beta)$ and shape given by ${\bf A}_\beta$.  A distance function can be defined that measures the number of standard deviations that the endpoint of a trajectory is away from the final wave packet centroid.  It is given by
\begin{equation}
f_\beta(\vec p_0,\vec q_0;t) = (\delta \vec p_t, \delta \vec q_t) \cdot {\bf A}_\beta \cdot (\delta \vec p_t, \delta \vec q_t)
\end{equation}
where $(\delta \vec p_t,\delta \vec q_t)  = (\vec p_t - \vec p_\beta, \vec q_t - \vec q_\beta)$.  The trajectory endpoint $(\vec p_t, \vec q_t)$ clearly is a function of the initial conditions.  As $(\vec p_0, \vec q_0)$ is varied, each isolated minimum corresponds to a unique classical pathway.  However, these minima come in one parameter families with time and one only needs the local minima in time on the central energy surface, as previously discussed.

An excellent predictor of how much a saddle can contribute to a correlation function is given by this distance function.  Considering the sum of the initial and final distances of a seed trajectory, $\gamma$,
\begin{equation}
D_\gamma= f_\alpha(\vec p^\gamma_0,\vec q^\gamma_0;0) + f_\beta(\vec p^\gamma_0,\vec q^\gamma_0;t)
\end{equation}
Thus, all the hard work of finding complex saddle trajectories is reduced to finding the local minima of $D_\gamma$ in the reduced dimensional space determined by the properties of ${\bf A}^{-1}_\alpha (\tau)$ and ${\bf M}_\tau^{-1}$.

For the seed trajectories identified by the minima of $D_\gamma$, the function ${\rm e}^{-D_\gamma}$ gives a fairly good rough estimate of the suppression of the semiclassical contribution due to the associated saddle trajectory's mismatch with the real centroids of the initial and final wave packets.  Therefore, the matrices ${\bf A}_\alpha$ and ${\bf A}_\beta$ can be used to cut off the search space domains.  Typically it is found that the chirp given by a saddle family's contribution tends to be rather insignificant if $D_\gamma\ge 10$.  Notice that this provides a cut off criterion that does not increase with $N$ increasing.  This has the consequence that for each single degree of freedom, the phase space coordinate of a relevant saddle trajectory tends to get closer to the central trajectory as $N$ increases. 

\subsection{The role of symmetry}

The symmetries of a quantum Hamiltonian lead to an important role for the irreducible representations of the associated groups with respect to the properties of the eigenvalues and eigenfunctions.  The Hilbert space can be represented by a basis which separates into subspaces, each having specific transformation properties with respect to the actions of the associated group operators.  If an initial state respects at least some part of the dynamical or fundamental symmetries of the system, it necessarily can be constructed from a subspace of the full Hilbert space.  Quantities, such as the autocorrelation function, Eq.~(\ref{ac}), must have enhanced long time averages as a result.  The enhancement depends on the ratio of the full Hilbert space dimensionality relative to the appropriate subspace.

This, of course, must be reflected in the  semiclassical theory.  The dynamical effects are accounted for by the transformation properties of the saddle trajectories.  It suffices to consider the saddle trajectories' initial conditions and how they transform under the group operations.  If an operation returns the same initial condition, no multiplicity is implied, otherwise there must be a replica of the saddle trajectory given by the particular operation.  Hence the rule, highly symmetric saddle trajectory initial conditions leads to low multiplicities, and low symmetry initial conditions leads to higher multiplicities. 

Depending on the Hamiltonian and initial and final states then, there will be a symmetry reduced fundamental domain in the phase space, which can be used to search for saddles.  The saddles within the other domains follow by a symmetry operation.  The precise domain boundaries depend on the subspace, i,e.~set of necessary search directions,
\begin{equation}
\left\{\left( \begin{array}{c} \delta \vec p_0 \\ \delta \vec q_0 \end{array} \right)_j \right\}\ .
\end{equation}
They collectively define a volume, which can be decomposed into fundamental domains.  

This imposes a certain structure on the eigenvectors giving the search directions.  If the search domain is comprised of a single eigenvector, hence the eigenvalue is non-degenerate, the symmetry operation applied to the vector has to return the negative of the vector.  The symmetry imposed eigenvector structure in such a case is immediately visible with a cursory glance.  However, in higher dimensional search spaces and especially if there are search directions associated with degenerate eigenvalues (equal stretching rates, $\lambda_{j,+}$), it may happen (as seems rather likely) that the structure of the eigenvectors is somewhat hidden from view and it can be rather difficult to identify fundamental domain boundaries.  In such a case, a rotation of the degenerate search directions can aid immensely their identification and the structure imposed on the eigenvectors.  A non-trivial example is shown in Sect.~\ref{6s} where there is a $6$-fold symmetry in a $2$-dimensional space, but the fundamental domain cannot be selected as just any $60^\circ$ wedge in the plane.  The eigenvectors that emerge from the stability analysis have to be rotated to identify the boundaries.  Once the analysis is completed and the boundaries are properly identified though, the reduced domain can be used to accelerate the saddle search and the construction of a semiclassical approximation.

As a final remark, note that there are significant symmetry effects on the dynamics of multidimensional quantum systems, which semiclassical theory is entirely capable of addressing in detail.  In particular, they affect the far-out-of-equilibrium dynamics of a many-body system such as represented by the Bose-Hubbard model discussed in Sect.~\ref{bhms}, some of which is addressed in detail there.

\section{Bose-Hubbard model saddle trajectories}
\label{bhms}

In recent studies calculating post-Ehrenfest quantum many-body interferences~\cite{Tomsovic18b,Ullmo18} and coherence effects~\cite{Schlagheck18}, this method was used to find saddle trajectories for a Bose-Hubbard model in a ring configuration.  The quantum Hamiltonian contains tunable nearest neighbor hopping and two-body interaction terms, and can be expressed as
\begin{equation}
\label{bhm}
\hat H = -J \sum_{j=1}^N \left(a^\dagger_j a_{j+1} + h.c.\right) +
\frac{U}{2} \sum_{j=1}^N \hat n_j \left(\hat n_j - 1 \right) 
\end{equation}
where $N$ is the number of sites in the ring and determines the number of degrees of freedom.  $U$ is a measure of the strength of the two-body interaction, which depends on the s-wave scattering length.  $J$ controls the tunneling amplitude, which depends on the well depth.  There are two constants of the motion, the energy and total number of particles, $\hat n_T=\sum_j \hat n_j$.

A mean field analysis~\cite{Pitaevskii03,Castin98} leads to a corresponding classical Hamiltonian, which follows from the introduction of the quadrature operators $(\hat q_j, \hat p_j)$ defined as 
\begin{align*}
\hat a_j         &= \frac{\hat q_j + i \hat p_j}{\sqrt{2}} \\
\hat a_j^\dagger &= \frac{\hat q_j - i \hat p_j}{\sqrt{2}} \; .
\end{align*}
and subsequent replacement by $c$-numbers.  After accounting for operator ordering issues, this gives
\begin{eqnarray}
\label{hamiltonian}
H_{cl} &=& -J \sum_{j=1}^N q_j q_{j+1} + p_j p_{j+1} + \frac{U}{2}
\sum_{j=1}^N \left(\frac{q_j^2 + p_j^2}{2}\right)^2 \nonumber \\ 
&& - U \sum_{j=1}^N \frac{q_j^2 + p_j^2}{2}
\end{eqnarray}
It is a quartic function of the phase space variables and straightforwardly analytically continued to complex variables.  The second constant of the motion is given by
\begin{equation}
n_{cl} = \sum_{j=1}^N \frac{q_j^2 + p_j^2}{2}
\end{equation}
and is the fixed total number of particles for a classical trajectory.

\subsection{Quantum and classical symmetries}
\label{qcsym}

This Bose-Hubbard model, Eq.~(\ref{bhm}), has the following discrete symmetries: cyclic permutation, reverse index ordering (clockwise/counterclockwise ring), and time reversal invariance.  The first two symmetries lead to groups of order $g_N=2N$ for $N\ge 3$.  For $N\le2$, reversing the index ordering is identical to cyclic permutation and hence $g_N=N$ for $N=1,2$.  

Furthermore, the model has a continuous symmetry, $U(1)$, in which multiplying the set $\{\hat a_j\}$ by a phase ${\rm e}^{i\theta}$, and hence the $\{\hat a^\dagger_j\}$ by the complex conjugate phase leaves the Hamiltonian invariant.  This is equivalent to the rotation of the quadrature operators,
\begin{equation}
\label{quadrot}
\left( \begin{array}{c}
\hat p_j^\prime \\
\hat q_j^\prime
\end{array} \right) = \left( \begin{array}{cc}
\cos \theta & -\sin \theta \\
\sin \theta & \cos \theta \end{array} \right) \left( \begin{array}{c}
\hat p_j \\
\hat q_j
\end{array} \right)
\end{equation}
and similarly for the $c$-numbers.  Thus, all these symmetries are reflected in the classical dynamics and the initial conditions of the saddle trajectories.

From a theoretical perspective, one can design a symmetry group of interest quite easily for a many-body system of a type akin to the Bose-Hubbard model of Eq.~(\ref{bhm}).  For example, if hopping connects all the sites equally, the maximum discrete group of the Hamiltonian would be the permutation group (the symmetry group), $S_N$.  The actual constructive interference and long-time average enhancement factors would depend on the symmetry properties of the initial and final states.

\subsection{Coherent state density waves}
\label{csdw}

A coherent state density wave is a useful initial state for our demonstration purposes~\cite{Tomsovic18b,Ullmo18}.  Denote it
\begin{equation}
|{\bf n}\rangle = \prod_{j=1}^N \exp
\left(-\frac{\left|b_j\right|^2}{2} + b_j  a^\dagger_j \right)|{\bf 
0}\rangle 
\end{equation}
where each site $j$ of the ring potentially has a different mean number of particles $n_j=\left|b_j\right|^2$.  A coherent state density wave is populated as follows $|n,0,n,0,...,n,0\rangle$, where $n$ represents the {\bf mean} number of particles on that site (not to be confused with a Fock state density wave).  This notation is incomplete in that the phase of each $b_j$ is not specified.  Thus, we assume that the $\{b_j\}$ are all chosen real and positive if not indicated otherwise.  An example of an initial state that does have alternating phases of the $\{b_j\}$ is discussed near the end of Sect.~\ref{regime}.

In a configuration representation, initial coherent states appear as Gaussian wave packets
\begin{eqnarray}
\phi_\alpha(\vec x) &=& \pi^{-N/4} \exp\left[- \frac{\left(\vec x- \vec q_\alpha\right)^2}{2}  + i \vec p_\alpha \cdot \left(\vec x - \vec q_\alpha\right) \right. \nonumber \\
&& \left. + i \frac{\vec p_\alpha \cdot \vec q_\alpha}{2}\right]
\end{eqnarray}
where
\begin{equation}
\label{rotqp}
\sqrt{2}\ \vec b = \vec q_\alpha + i \vec p_\alpha
\end{equation}
This is in the form of Eq.~(\ref{wavepacket}) as it must be, except with a different phase convention given by the last term of the equation.  Its Wigner transform is
\begin{equation}
{\cal W}(\vec q,\vec p) = \pi^{-N} \exp\left[- \frac{\left(\vec q- \vec q_\alpha\right)^2}{2}  -  \frac{\left(\vec p- \vec p_\alpha\right)^2}{2}\right] 
\end{equation}
These equations define the wave packet centroid, phase convention, shape matrices ${\bf b}_\alpha =\mathbb{1}, {\bf A}_\alpha ={\bf A}_\beta =\mathbb{1}$, and $\hbar=1$.  If the components of $\vec b$ are chosen real and positive, there is just a shift in the position centroids per site,
\begin{equation}
\phi_\alpha(\vec x) = \pi^{-N/4} \exp\left[-  \sum_{j=1}^N  \frac{\left(x_j-\sqrt{2n_j}\right)^2}{2} \right] 
\end{equation}
This gives rise to the corresponding density operator Wigner transforms,
\begin{equation}
{\cal W}(\vec q,\vec p) = \pi^{-N} \prod_{j=1}^N
\exp \left[ -\left(q_j-\sqrt{2n_j}\right)^2 - p_j^2 \right] 
\end{equation}

\subsection{4-site coherent state density wave}
\label{4s}

Consider a 4-site ring with initial coherent state density wave $|20,0,20,0\rangle$ ($b_j$ chosen real and positive), and let the interaction strength be $U=0.5$.  There are two time scales in the dynamics for the Bose-Hubbard model without hopping given by 
\begin{equation}
\tau_1 = \frac{2\pi}{U n_j} = 0.63 \qquad \tau_2 = \frac{2\pi}{U} =
4\pi= 12.57 \; ,
\end{equation}
$\tau_1$ is a classical scale associated with first return of classical trajectories, and $\tau_2$ is a quantum scale associated with the revival of the initial quantum state~\cite{Greiner02b}.  

We fix the hopping strength to be $J \! =\! 0.2$, which perturbs the dynamics, but leaves the system in the strong interaction regime.  The autocorrelation function, Eq.~(\ref{ac}), constructed semiclassically using these saddles is pictured in upper panel of Fig.~1 of Ref.~\cite{Tomsovic18b}, where it is seen to have quite a complicated set of oscillations.  The revivals and fractional revivals $(1/2,1/3,...)$ are reduced, but still visible; the semiclassical saddle point formulas can be found there and are not repeated here.  They require a great deal of delicately balanced quantum interference to reconstruct, but the semiclassical approximation does so.  This could only happen if one has identified all or nearly all of the contributing saddles.

\subsubsection{Search directions}
\label{sd}

The first step is to construct and diagonalize the matrix ${\bf M}_\tau\cdot{\bf M}_\tau^T$ (since ${\bf A}_\alpha = \mathbb{1}$) of the initial condition $(\vec p_\alpha,\vec q_\alpha)$ [wave packet centroid] for a long enough propagation time that the eigenvectors have converged to their asymptotic directions; a value of $\tau$ on the order of $(1.5-2) \times \tau_2$ was sufficiently asymptotic.  Of the $4$ eigenvalues greater than unity, one dominates, is at least somewhat exponentially unstable, and is given by $5.8 \times 10^{11}$ at $t=16$.  The next largest eigenvalue is $1.4 \times 10^{5}$ ($10^6$ times smaller), and its eigenvector is associated with the direction of maximal change in energy, which as pointed out in subsection~\ref{distinguish} is not necessary to search.  The final $2$ eigenvalues are nearly degenerate with value $1.8$, and are entirely irrelevant.  Thus, this case can be reduced to a $1$-parameter search for saddles without losing any dynamical information on the time scale of $1-2$ revivals, say, less than $2\tau_2$; the straightforward search dimensionality for this case would have required $8$ parameters.  The eigenvector associated with the largest eigenvalue is used in conjunction with Eq.~(\ref{delta2}) to determine the sole line of initial conditions necessary to search for saddles.

\subsubsection{Saddles}
\label{sadd}

\begin{figure}[tbh]
\includegraphics[width=8.7 cm]{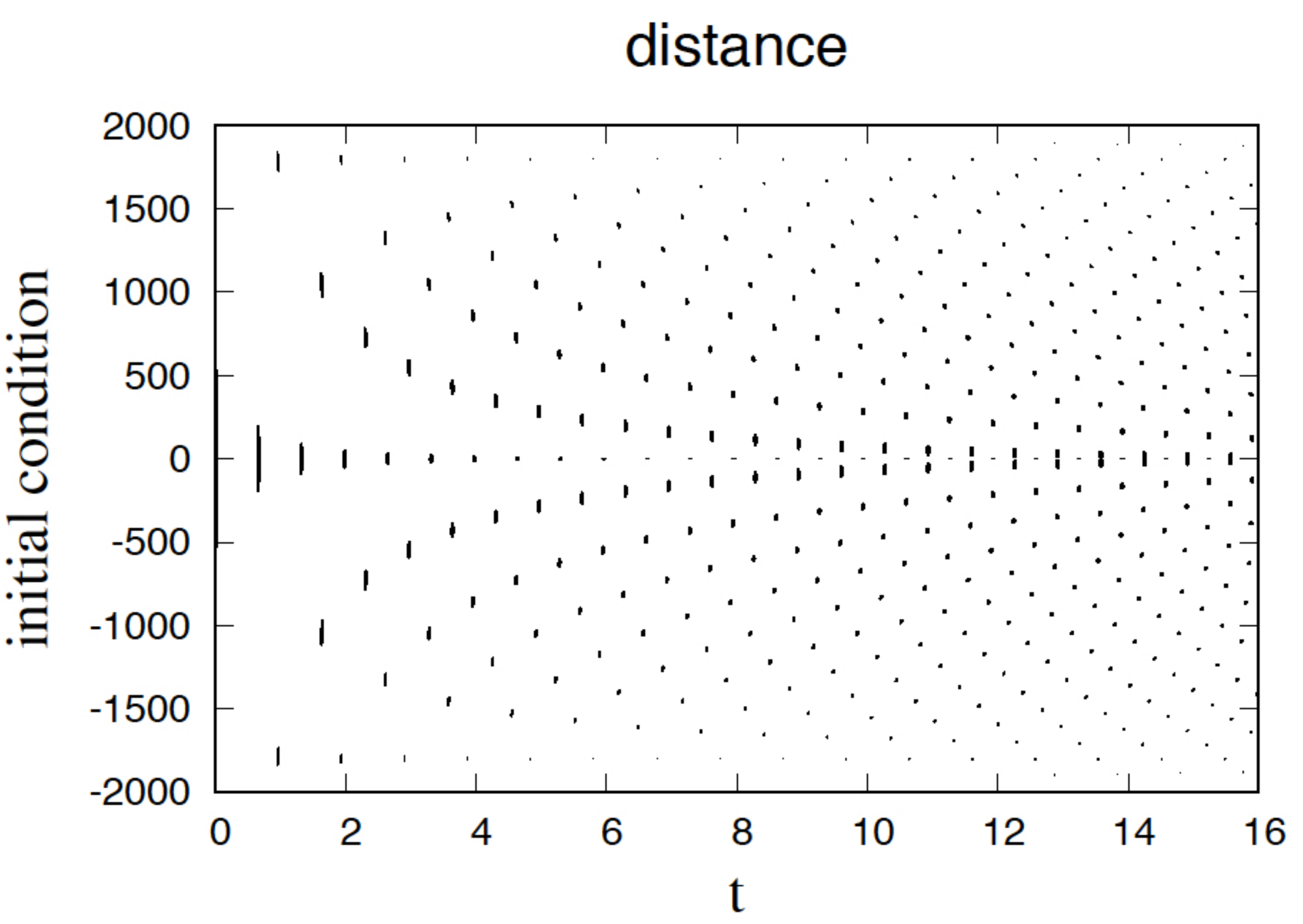}
\caption{View of seed trajectory locations in time and initial conditions.  For $4000$ initial conditions the distance function is calculated as a function of time.  The initial conditions are chosen uniformly along the eigenvector mentioned in the text across the interval $[-4\sigma,4 \sigma]$ corresponding to the Wigner transform of the coherent state density wave.  The points are blackened where $D_\gamma \le 20$.  The initial conditions are labeled by an index on the $y$-axis.  The full discrete symmetry is encapsulated by a reflection symmetry with respect to the $x$-axis.  The multiplicity $1$ saddles are found using the $y=0$ line seed trajectories, and the rest have multiplicity $2$. \label{fig2}} 
 \end{figure}

In Fig.~\ref{fig2}, the points where the distance function, $D_\gamma\le 20$, are blackened and plotted as a function of time and initial 
conditions.  The search direction is given by the vector $\left( \begin{array}{c} \delta \vec p_0 \\ \delta \vec q_0 \end{array} \right)_1$ of Eq.~(\ref{delta2}).  The propagation time and initial conditions of each seed trajectory are selected by the one whose distance is minimized within each isolated blackened region.   

\begin{figure}[tbh]
\includegraphics[width=8.5 cm]{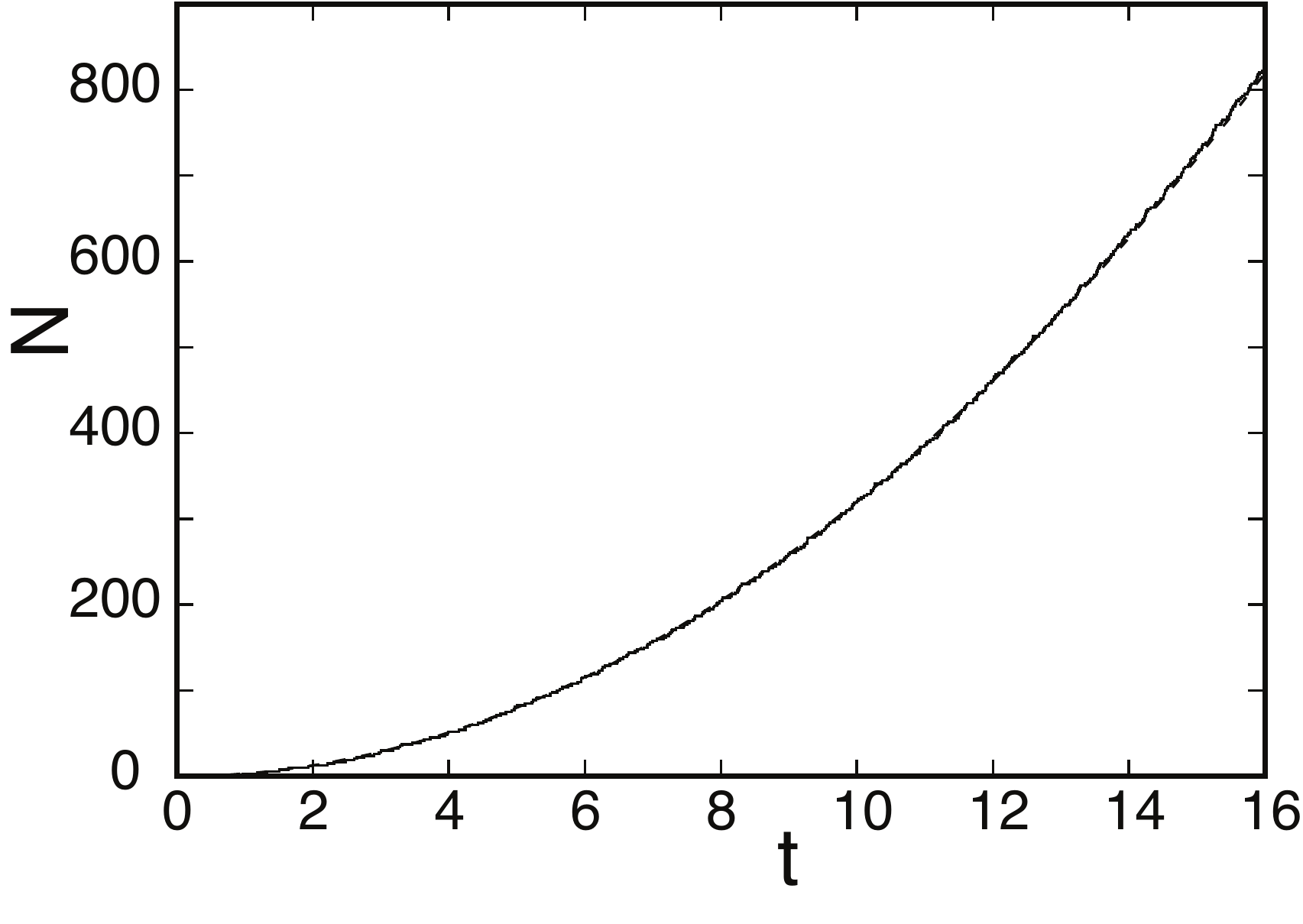} 
\caption{Total number of saddles as a function of time.  The saddles counted are those found with seed trajectories with a $D_\gamma \le 20$.  A dashed quadratic curve is shown as a guide.   \label{fig3}} 
 \end{figure}

After using the Newton-Raphson scheme, each region leads to a unique saddle trajectory family with a semiclassical contribution to ${\cal A}(t)$ similar to the one shown in Fig.~\ref{fig1}.
The total number of saddles found as a function of time is illustrated in Fig.~\ref{fig3}.  
In this particular dynamical case, one can be fairly certain that all the saddle trajectory families have been found (up to a certain significance) due to the highly structured locations of the distance minima.  Despite the high degree of instability in the largest eigenvalue, the saddle number is increasing similarly to that of a system in a near-integrable dynamical regime, i.e.~a linearly increasing density of saddles in time leads to a pure quadratic total count of saddles up to some fixed time.  If the system were behaving as a purely chaotic dynamical system, the number of saddles found would increase as an exponential function.

In addition, it is possible to see an approaching problem as time increases.  Above and below the central horizontal axis $(0$-line) are regions approaching each other in pairs, which implies a coalescence of saddle points once they overlap.   Beyond a certain time, to avoid singularities in the semiclassical theory the coalescing saddles will require a uniformized approximation of the kind discussed in~\cite{Chester57}.  

If one uses the second largest eigenvector, $\left( \begin{array}{c} \delta \vec p_0 \\ \delta \vec q_0 \end{array} \right)_2$, one finds only the symmetric saddles that can be found using the single trajectory with initial condition $(\vec p_\alpha,\vec q_\alpha)$.  As this vector is associated with the normal to the \begin{figure}[tbh]
\includegraphics[angle=-90,width=8.5 cm]{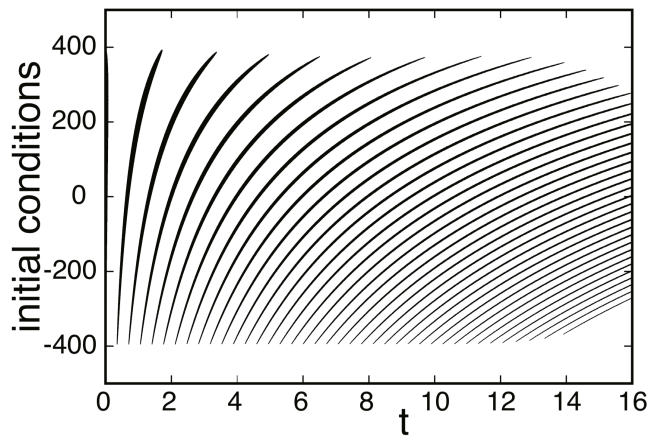}\vskip -.5cm
\caption{Equivalent of Fig.~\ref{fig2} for the eigenvector perpendicular to the energy surface.  This eigenvector is associated with the second largest eigenvalue.  For $1000$ initial conditions the distance function is calculated as a function of time.  The initial conditions are chosen uniformly along the eigenvector across the interval $[-4\sigma,4 \sigma]$ corresponding to the Wigner transform of the coherent state density wave.  The points are blackened where $D_\gamma \le 20$.  The initial conditions are labeled by an index on the y-axis. \label{fig4}} 
 \end{figure}
energy surface, this direction preserves the symmetries of the central trajectory, and it does not lead to any new saddles beyond those found with the central trajectory, rather it just generates the families of each of the saddles which are fully symmetric.  

It is interesting however to illustrate this point.  This vector's equivalent of Fig.~\ref{fig2} is shown in Fig.~\ref{fig4}.  In the strong interaction regime, there is strong shearing perpendicular to the energy surface and this makes each saddle trajectory family contribute over a wide range in time; recall Fig.~\ref{fig1}.  It is possible to deduce from this figure the width of semiclassically-contributing time window for any particular symmetric saddle.  Select one of the contiguous blackened regions, and fix any time that intersects it.  There will be a minimum distance trajectory for that fixed time, somewhere near the middle of the fixed time vertical line's intersection with the region.  It can be used to locate some particular symmetric saddle.  If one differentially shifts the time back and forth enough to intersect the entire chosen region, the continuous collection of saddles forms a saddle family exactly as discussed in Sects.~\ref{family},\ref{distinguish}.  In fact, one could construct a saddle family this way with a large number of real seed trajectories, one for each fixed time, but the method discussed in Sects.~\ref{family},\ref{distinguish} is much more reliable and faster.  It is better not to use this direction in the saddle searches as mentioned earlier.  

The time interval that intersects the chosen region is the contributing time window of a saddle family, just as pictured in Fig.~\ref{fig1}.  Therefore, the regions further to the right (increasing time), corresponding to later arriving saddle families, which are more horizontally tilted, have corresponding saddle families that contribute to the autocorrelation function over wider time windows.  It suffices to project any particular region seen in Fig.~\ref{fig4} onto the time axis to read off the width of that saddle family's contribution in time.

\subsubsection{Symmetries}
\label{symm}

The initial condition associated with the coherent state density wave centroid, $\vec p_\alpha = \vec 0$ and $\vec q_\alpha =  (20, 0, 20, 0)$, is invariant under some of the symmetry operations that leaves the Bose-Hubbard model invariant, i.e.~a double hop cyclic permutation and time reversal invariance.  These symmetries have a number of consequences for the autocorrelation function defined in Eq.~(\ref{ac}).  

Two consequences are handled quickly.  First, time reversal invariance ensures that ${\cal A}(-t) = {\cal A}^*(t)$, but does not otherwise lead to symmetry related saddles (multiplicity greater than 1) forward in time.  Second, any choice of rotation via Eq.~(\ref{quadrot}) acting on the variables of $(\vec p_\alpha, \vec q_\alpha)$ of Eq.~(\ref{rotqp}) leaves the autocorrelation function invariant.  This is reflected in a symmetry of the classical trajectories, whereby a rotation of initial conditions of this sort leads to a trajectory linked to the former by rotation.

The remaining cyclic permutation and index reversal symmetry does lead to symmetry related saddles and this is visible in the symmetry of Fig.~\ref{fig2}.  For this case, a saddle may be unique or duplicated elsewhere in phase space by the double cyclic permutation.  To be unique, the initial condition of the saddle trajectory must have the same symmetry as $(\vec p_\alpha;\vec q_\alpha)$.  In other words, if the initial condition position is the same for sites $1$ $\&$ $3$ (the full symmetry is there, but observing just those two site positions identifies it), it has multiplicity $1$, and if they are different, then it has multiplicity $2$.  All of the initial conditions in the neighborhood of $(\vec p_\alpha;\vec q_\alpha)$ have lower symmetry than it does (excluding the direction of maximal change in energy).  Thus, the only multiplicity $1$ saddles arise in the Newton-Raphson search from seed trajectories found using the initial condition $(\vec p_\alpha,\vec q_\alpha)$; this is effectively a zero parameter search of initial conditions.  The regions straddling the central horizontal axis  have multiplicity $1$.  The rest of the saddles have multiplicity $2$ and in this case arise from a one parameter search.  In fact, one can reduce the search regime to the region above the central axis and multiply the contributions of the saddles to ${\cal A}(t)$ by their multiplicity index.

Since total particle number is a conserved quantity, and this example is in the strong interaction regime, the structure of the optimal vector search direction can be understood by simple arguments.  For the moment, assume the hopping is turned off, and the classical dynamics are quasi-periodic.  In order for a trajectory to return close to its initial conditions, as must be the case for an autocorrelation function, the period of motion for each site must be nearly integer multiples of each other.  The shearing is strongest perpendicular to the energy surface for each site.  Also, there is almost no frequency change for the unoccupied orbitals.  Since for the coherent state density wave chosen, the periods of motion for site 1 \& 3 are identical, the strongest change in their period ratio away from unity, while preserving the total particle number, is for either site 1 to increase its occupancy and site 3 to decrease by the same or vice versa.  Furthermore, with $b_j$ real and positive, the perpendicular to the energy surface involves only $q_1$ or $q_3$, no momenta (the perpendicular vector at a point on a circle lies along the continuation of the radial line from the center to that point).  Thus, the search direction incorporates vanishing changes in momenta, and a change in position $\delta \vec q = (\delta q, 0, -\delta q, 0)$.  Even after turning the hopping term back on, the direction is dominated by these changes.  It is clear why Fig.~\ref{fig2} has a reflection symmetry with respect to the central axis, $(\delta q, 0, -\delta q, 0)$ and $(-\delta q, 0, \delta q, 0)$ are related by double cyclic permutation or index reversal (with a shift).  As the story gets more complicated for greater numbers of sites, we introduce a shorthand for this search direction $(\delta q, 0, -\delta q, 0) \equiv (\delta n, -\delta n)$, ignoring the unoccupied sites or the difference between position and momentum; note that in this shorthand, the second direction, the one associated with the perpendicular to the energy surface and Fig.~\ref{fig4}, is denoted $(\delta n, \delta n)$.  

One implication of the irrelevance of the unoccupied sites and momentum generally in the search directions is that in the strong interaction regime, it is never necessary to search more than $N/2-1$ dimensional spaces to find all the contributing complex saddles up to intermediate time scales.  

\subsection{6-site coherent state density wave}
\label{6s}

Consider next a 6-site ring with initial coherent state density wave $|10,0,10,0,10,0\rangle$, and let the interaction and hopping strengths, respectively, be $U=1.0$ and $J=0.2$.  In this case, the largest eigenvalue is doubly degenerate, and the initial condition search directions correspond very roughly to $(\delta n, -\delta n, 0)$ and $(\delta n, \delta n, -2\delta n)$; the normalization is not given by the notation.  \begin{figure}[tbh]
\vskip .3 cm\includegraphics[angle=-90,width=9 cm]{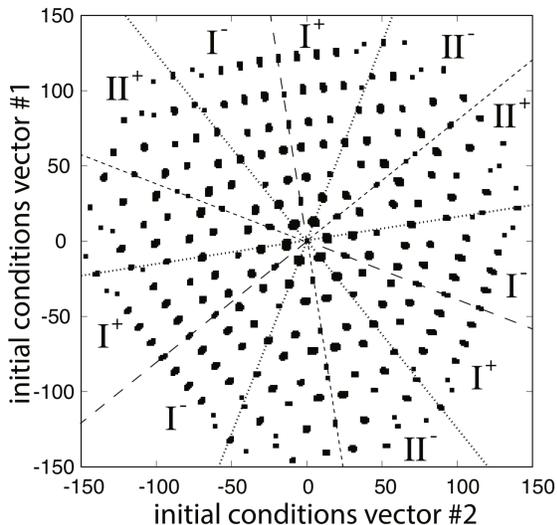}
\caption{$D_\gamma$ near the revival time.  Each black spot gives rise to a single seed trajectory, and hence there is a one-to-one correspondence between spots and saddles.  A $300 \times 300$ grid of initial conditions were used to calculate $D_\gamma\le 20$ at $\tau_2$.  The $6$ dotted lines correspond to the $6$ symmetry related vectors of $(\delta n, -\delta n, 0)$.   The $3$ long dashed and $3$ medium dashed lines correspond to cyclic permuations of $(\delta n, \delta n, -2\delta n)$ and its negative.  They separate the plane into twelve domains.  The domains $I^-$ and $I^+$ are mirror images of each other about the dashed line between them, similarly for $II^-$ and $II^+$.  The $3$ copies of each domain, $I^\pm$ and $II^\pm$, are related by $120^\circ$ rotation.  One choice for a fundamental domain would be the sum of the regions $I^-$ and $II^+$ adjacent on the right side of the figure.  \label{fig5}} 
 \end{figure}
The next largest eigenvector corresponds to the energy surface perpendicular, $(\delta n, \delta n, \delta n)$.  No other search directions are even remotely relevant.  This information was used to locate the roughly $5000$ saddle families up to $t=12$.

The equivalent of Fig.~\ref{fig2} would be 3-dimensional.  Instead, Fig.~\ref{fig5} shows where  
$D_\gamma \le 20.0$ in the plane of initial condition search directions, but showing a cross-section by fixing the time to be equal to $\tau_2$ ($=2\pi$).  The double degeneracy turns out to be necessary to accommodate the higher symmetry.  For example, form the vector from the sum of the two above.  The resulting vector is equivalent to $(\delta n, 0, -\delta n)$, which is an odd permutation of the first vector.  The difference gives a positive permutation.  In fact, using appropriate normalization and summing or subtracting (recall that ${\bf A}_\alpha=\mathbb{1}$), it is possible to construct in the plane of initial conditions all $6$ symmetry related versions of $(\delta n, -\delta n, 0)$, uniformly spread out with $60^\circ$ between them.  Similarly, it is possible to build the $3$ cyclic permutations of $(\delta n, \delta n, -2\delta n)$, as well as the $3$ cyclic permutations of the negative $(-\delta n, -\delta n, 2\delta n)$.  Unlike $(\delta n, -\delta n, 0)$, which gives rise to a multiplicity of $6$, these two sets cannot be mapped onto each other by a symmetry operation and saddles associated with them only come in multiplicities of $3$.  These twelve lines are indicated in Fig.~\ref{fig5}.  They separate the fundamental domains, which can be mapped onto each other by either a cyclic permutation or index reversal.  Thus, to find all the saddles, it is only necessary to search in $1/6^{th}$ of the initial condition plane.  For example, the 
$60^\circ$ wedge encompassing areas $I^-$ and $II^+$ on the right hand side would give a complete set of saddles.  Those emanating from the central initial condition have multiplicity $1$, those on the 
two symmetry lines at the bottom of $I^-$ and the top of $II^+$ have multiplicity $3$, and the rest have multiplicity $6$.

\begin{figure}[tbh]
\includegraphics[width=8 cm]{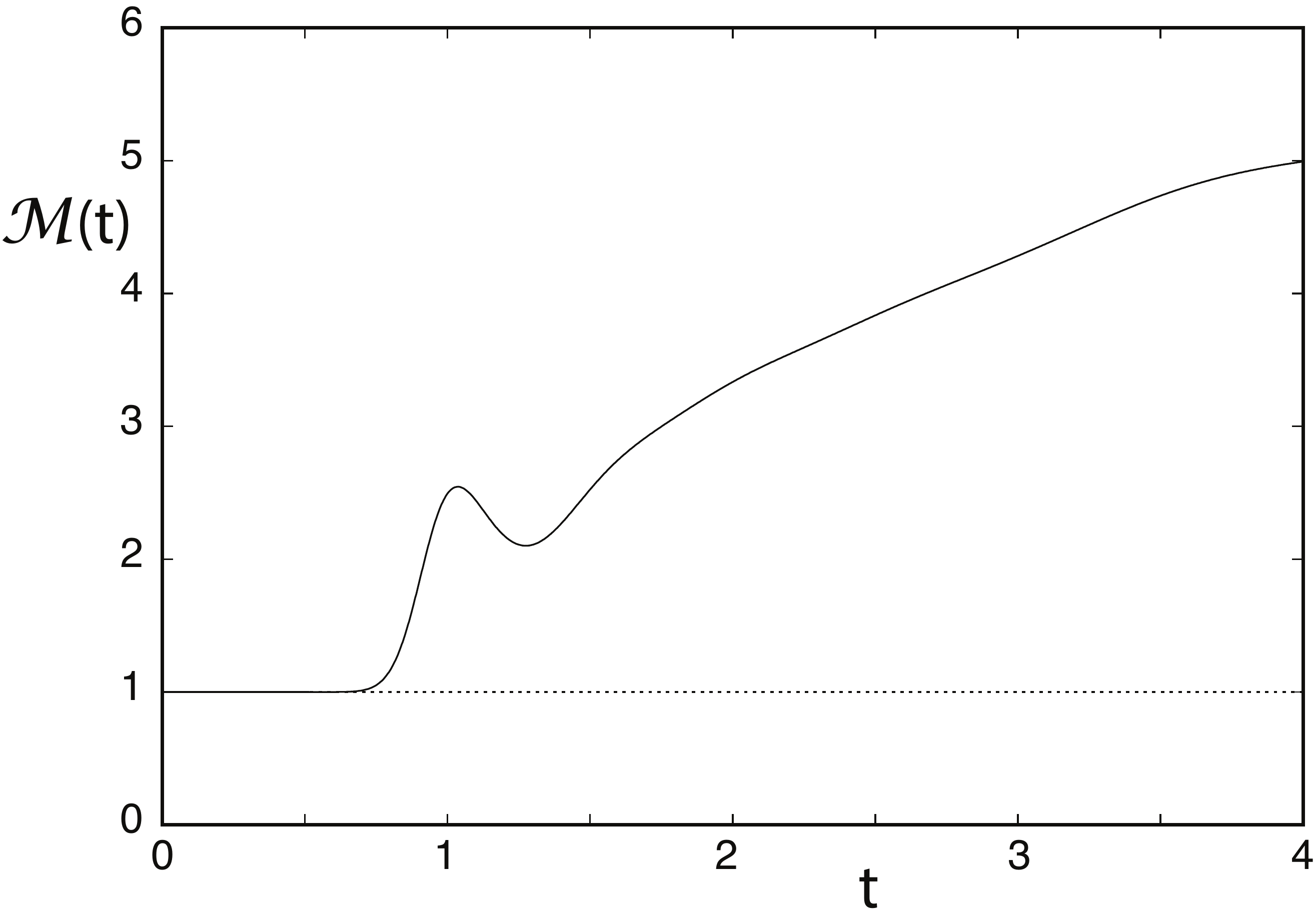}
\caption{$\cal M$ as a function of time.  The effective multiplicity of the saddles begins at $1$ and increases toward the maximum possible of $6$ in this case as time increases.  At long times, it saturates at $6$ and remains there as these saddles come from a higher dimensional space of initial conditions than the others.  \label{fig6}} 
 \end{figure}

For a highly symmetric point in phase space, typically it takes time for the neighboring lower symmetry trajectories to return.  This can be seen in this example by calculating an intensity waited average multiplicity of the saddles as a function of time.  It is given by
\begin{equation}
{\cal M}(t) = \frac{\sum_{j} g^2_j \left|{\cal A}_j(t)\right|^2}{\sum_{j} g_j \left|{\cal A}_j(t)\right|^2}
\end{equation}
where the index $j$ runs only over the symmetry reduced set of saddles.  Figure~\ref{fig6} illustrates the 
result for the $6$-site ring example.  The higher symmetry/lower multiplicity saddles dominate at short times and give way to dominance by the highest multiplicity saddles at longer times. 

\subsection{Remarks on the 8-site ring} 
\label{r8c}

The $8$-site model has a new feature, the 4 cyclic group has a 2 cyclic subgroup.  There will be saddles of degeneracies, $(1,2,4,8)$.  There are $3$ search directions necessary to construct the maximum $8$-fold saddle degeneracy, and a search for all the relevant saddles to long times will require a significant computational effort, but is quite possible to do.  Nevertheless, compared with the $16$-dimensions required of the straightforward search method, this is a great advance.  The fourth largest eigenvalue will be associated with the normal to the energy surface, $(\delta n, \delta n, \delta n, \delta n)$, and along with the remaining ones can be entirely ignored.

In greater detail, consider the case for $J=0.5$ and $U=0.5$ and an initial coherent state density wave $|40, 0, 40, 0, 40, 0, 40, 0\rangle$.  It turns out that the search direction associated with the most unstable eigenvector direction is $(\delta n, -\delta n, \delta n, -\delta n)$.  This direction can capture the saddles of multiplicity $2$; as usual multiplicity $1$ saddles require only the wave packet central orbit.  The two choices for the fundamental search domain are either the positive half line or the negative half line along this direction.  

It is slightly more complicated to determine the fundamental search domain for the multiplicity $4$ saddles.  The second and third most unstable directions are equally unstable (degenerate eigenvalues) and are roughly given by $(\delta n, 0, -\delta n, 0)$ and $(0, \delta n, 0, -\delta n)$.  Actually, due to the degeneracy, the two eigenvectors that emerge from the calculations are not these two, but rather a linear combination that hides the simple structure of these two vectors.  It is necessary to recognize that rotating the two calculated vectors generates the two above, which are then simpler and related by a cyclic permutation.  With the vectors above, four choices for a fundamental search domain could be given by the full line along $(\delta n, -\delta n, \delta n, -\delta n)$ and either the positive or negative half line along either $(\delta n, 0, -\delta n, 0)$ or $(0, \delta n, 0, -\delta n)$.  Another choice, though, could be the positive half lines of $(\delta n, -\delta n, \delta n, -\delta n)$ and $(\delta n, -\delta n, \delta n, -\delta n)$ plus the positive half lines of $(\delta n, -\delta n, \delta n, -\delta n)$ and $(0, \delta n, 0, -\delta n)$.  Some care must be exercised.  The choice of the positive half line along $(\delta n, -\delta n, \delta n, -\delta n)$ and the full line along $(\delta n, 0, -\delta n, 0)$ would turn out to miss half of the possible multiplicity $4$ saddles entirely (those found would come with a symmetry related partner).  A simple fundamental search domain for multiplicity $8$ saddles is the positive half lines of all three directions.

One curious feature is that is that the largest eigenvalue (a variance) turns out to be approximately $90$ times greater than the next $2$ eigenvector directions ($\sqrt{90}$ times more unstable).  This has some interesting consequences.  First, a crude guess would be that the earliest saddle of degeneracy $4$ should show up on a time scale roughly $\sqrt{90}$ times the first return time, $\tau_1$.  In fact, the first degeneracy-$4$ saddle appears at roughly $7.5\tau_1$.  Thus, there is a significant time separation of the initial appearance of saddles with multiplicities, $(1,2)$, relative to saddles with multiplicities, $(4,8)$.  The first return is non-degenerate, but by just after the second return, the quantum dynamics quickly becomes dominated by doubly degenerate saddles.  The situation remains this way until $7.5\tau_1$, when the first quadruply degenerate saddle arises.  They are few and weakly contributing, and so it still takes quite a bit more time for the quantum dynamics to be dominated by the highest degeneracy saddles.  If one is only interested in the initial interferences that arise in the dynamics, a $1$-dimensional search suffices for this $8$-site case, but to follow the dynamics long enough to see the emergence of the full symmetry enhancement in the autocorrelation function requires a full $3$-dimensional search.

\section{Identifying dynamical regimes using ${\bf M}_\tau \cdot {\bf A}_\alpha^{-1} \cdot {\bf M}_\tau^T$ and ${\bf M}_\tau^{-1}$}
\label{regime}

Many Hamiltonian systems depend on parameters, and they might be controllable in many cases, say for example, by varying externally controllable external field strengths.  For systems with many degrees of freedom, far out of equilibrium, it can be rather challenging to get a full understanding of the dynamics for an individual system, let alone for the range of dynamical possibilities of the system as a function of the parameters.  The analysis using the spectrum of ${\bf M}_\tau \cdot {\bf A}_\alpha^{-1} \cdot {\bf M}_\tau^T$ and its associated eigenfunctions (after mapping back with ${\bf M}_\tau^{-1}$) are ideally suited to elucidating the various dynamical regimes of such a system.  The results depend naturally on the phase space region of interest, which is determined by the central trajectory of the wave packet or coherent state.   For the Bose-Hubbard model of Sect.~\ref{bhms}, there are various transitions related to the relative strengths of the hopping ($J$-parameter) and interactions ($U$-parameter); one example is the much discussed superfluid-Mott insulator transition~\cite{Greiner02a}, another is the dynamical transition to more chaotic dynamics away from the pure hopping  and pure interaction limiting systems, which represent integrable systems~\cite{Kolovsky16}.  

To illustrate the idea, let $J=\cos\theta$ and $U=\sin\theta$ so that $J^2+U^2=1$.  There is a complete rotation of the system from pure hopping dynamics to pure interaction dynamics covered 
\begin{figure}[tbh]
\includegraphics[width=8.5 cm]{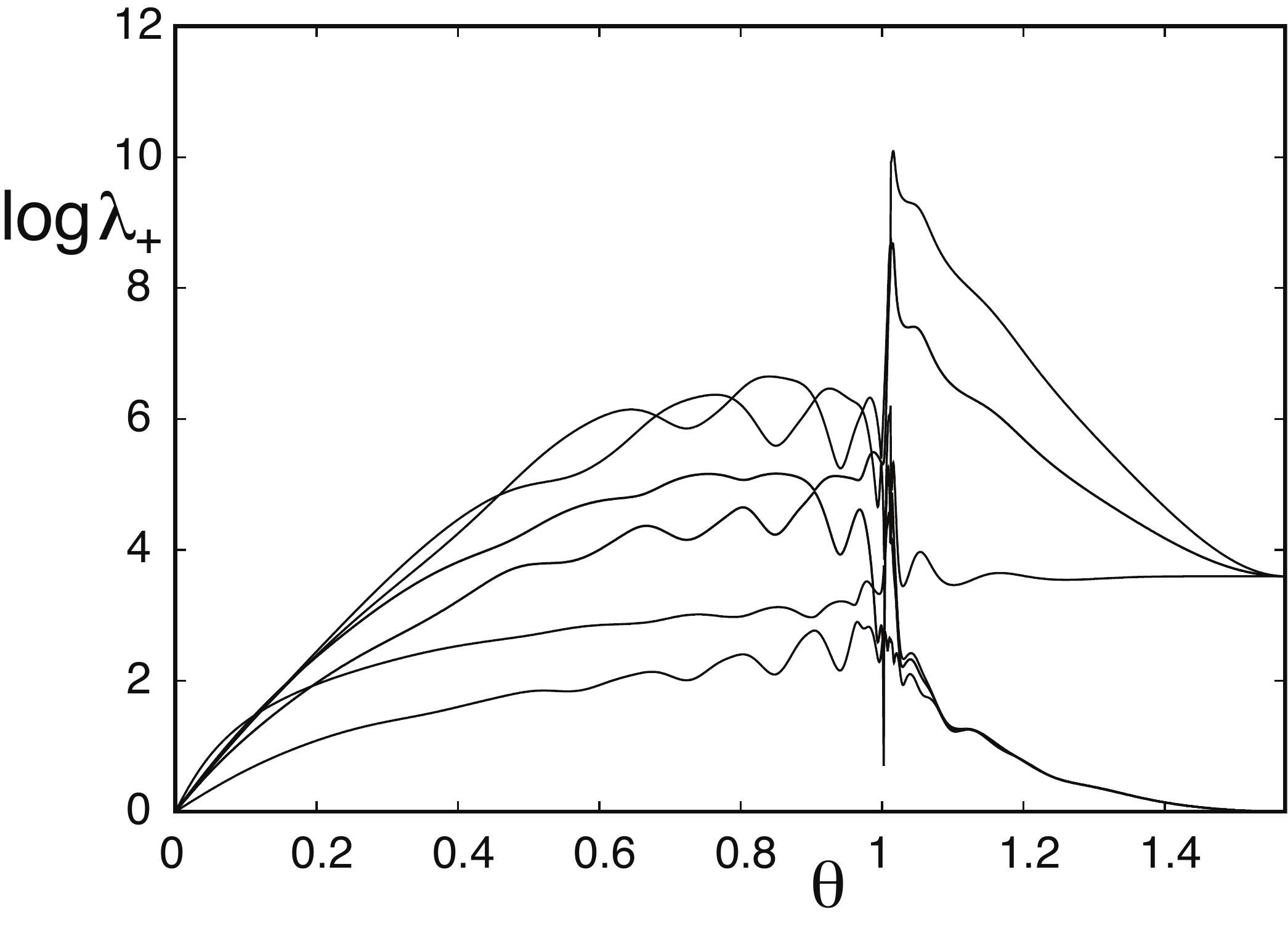}
\caption{Spectrum, $\{\log_{10}\left(\lambda_{j,+}\right)  \}$, of ${\bf M}_\tau \cdot {\bf A}_\alpha^{-1} \cdot {\bf M}_\tau^T$ as a function of $\theta$.  The initial state is a coherent state density wave for a ring with $8$-sites whose parameters are given in the main text.  All $8$ $\lambda_{j,+}$ are shown, but degeneracies make it appear as though fewer are plotted.  There is a strong realignment of the associated eigenvectors, $\left\{\left( \begin{array}{c} \delta \vec p_0 \\ \delta \vec q_0 \end{array} \right)_j\right\}$, at the transition point in the spectrum.  To the right of the transition, the eigenvectors essentially do not involve the initially unoccupied sites, whereas to the left, all the sites are involved and there is a double repetition around the ring in the structure of the eigenstates.  \label{fig7}}
\end{figure} 
by varying $\theta$ across the range $0\le \theta \le \pi/2$.  For an $8$-site ring, Fig.~\ref{fig7} shows the expanding part of the spectrum (the base $10$ logarithm of all eight $\lambda_{j,+}$) as a function of $\theta$ for a density wave coherent state with populated sites of mean number $n=5$ and $b=\sqrt{5}$ (i.e.~$|5,0,5,0,5,0,5,0\rangle$).  The spectrum is invariant with increasing particle number if the interaction strength $U$ is reduced by the increase.  Thus, for any value of the occupied sites, i.e.~coherent state density wave $|n,0,n,0,n,0,n,0\rangle$ generates the exact same $\theta$-dependent spectrum as Fig.~\ref{fig7} if one uses $U = \frac{5}{n}\sin\theta$ or rather $J^2+\left(\frac{n}{5}\right)^2 U^2=1$.

Moving from left to right, the spectrum exhibits a seemingly discontinuous change in the dynamical properties of the system near $\theta=1.01219704$ where the spectrum abruptly shifts and the eigenvectors completely rearrange their orientations.  This occurs at the same location independent of the number of sites in the ring.  \begin{figure}[tbh]
\includegraphics[width=8.5 cm]{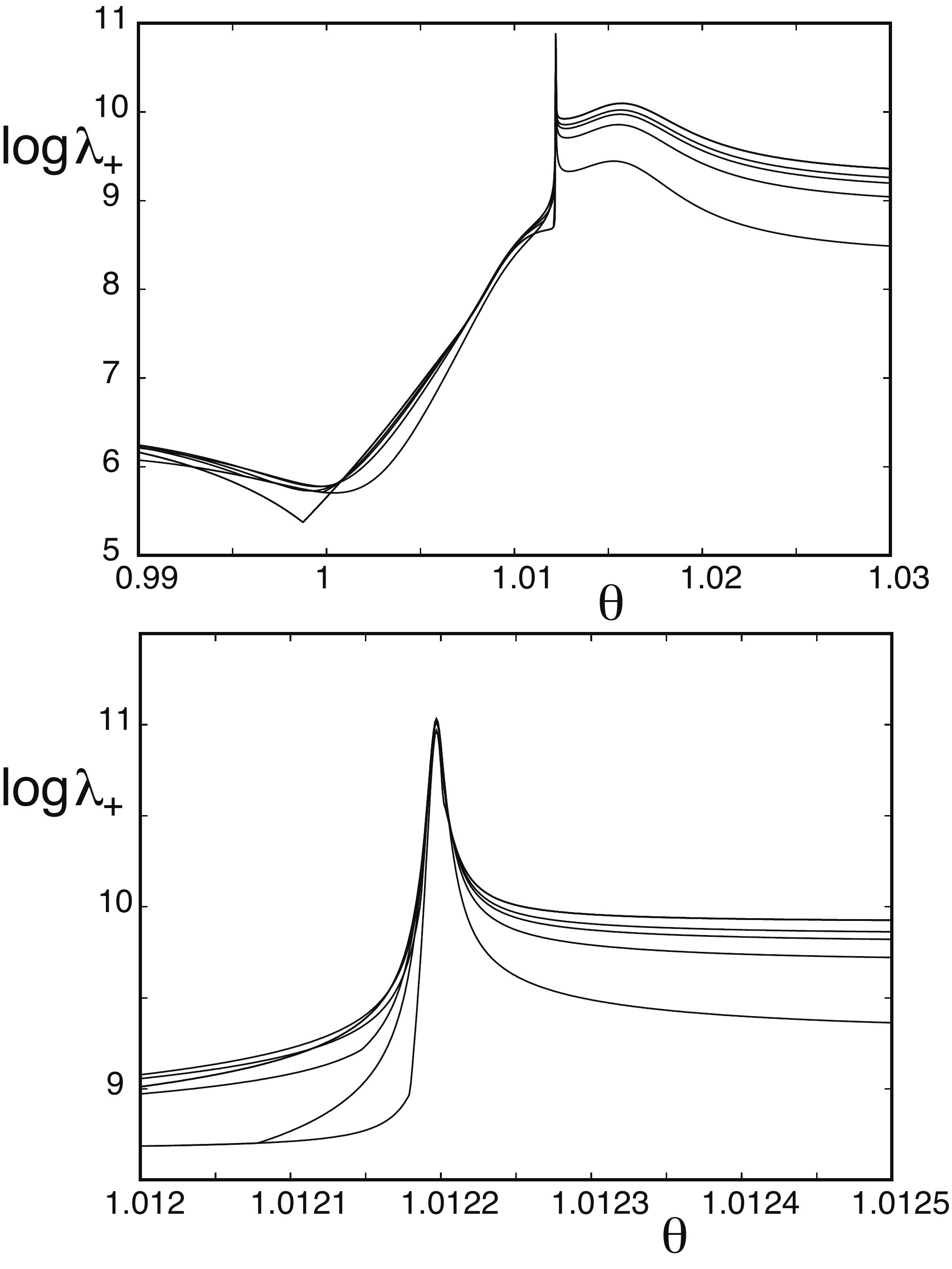}
\caption{Largest eigenvalue $\{\log_{10}\left(\lambda_{1,+}\right)  \}$, of ${\bf M}_\tau \cdot {\bf A}_\alpha^{-1} \cdot {\bf M}_\tau^T$ as a function of $\theta$.  The initial state is a coherent state density wave for rings with $4,6,8,10,12,14,16,18$-sites with populated sites of mean number $n=5$.  All $8$ cases of $\lambda_{1,+}$ are shown, but portions of the curves with fewer sites are copied in rings with greater numbers of sites.  For example, the $12$-site ring follows partly the $4$-site ring and partly the $6$-site ring results.  That makes it appear as though fewer examples are plotted.  The upper panel is a magnification of the transition region, which is magnified again in the lower panel near the sharp peak.  \label{fig8}}
\end{figure}
For example, Fig.~\ref{fig8} plots the largest eigenvector for all rings with even numbers of sites from $4$ to $18$ and $n=5$ as in Fig.~\ref{fig7}.  It turns out that although the transition is extremely abrupt, it is not discontinuous in either the number of sites or occupancies tending to infinity limits (assuming the appropriate scaling of $U$, and the peak occurs at a universal value of $nU/J=n\tan\theta\approx 8$ to an accuracy of better than one part in $10^6$.

To the right of this transition, the subspace ($4$-dimensional) of initial conditions involving the initially unoccupied sites rapidly fall towards zero, meaning that they have no involvement in the production of saddles.  Thus, the part of the initial conditions of saddle trajectories regarding those sites remain very nearly unoccupied for the entire time range that the semiclassical theory can be used to reconstruct the quantum dynamics.  The next larger eigenvalue is mostly horizontal on the right side and is related to the shearing perpendicular to the energy surface.  It has this general appearance in all the calculations regardless of site or particle numbers or initial conditions.  The fact that there are eigenvalues several orders of magnitude above it is an indicator of the presence of at least some chaotic dynamics in the system.  The next eigenvalue above is doubly degenerate and responsible for creating saddles that have multiplicities $4,8$ just discussed in greater detail in Sect.~\ref{r8c}.  The most unstable eigenvalue at the top is responsible for the multiplicity $2$ saddles.  The larger the gap between these two eigenvalues, the longer it takes for the effective saddle multiplicity, ${\cal M}(t)$, to transition from $1 \rightarrow 2 \rightarrow 4 \rightarrow 8$.

On the left side of the transition, the most unstable eigenvectors involve the initially unoccupied sites strongly.  They must satisfy the discrete symmetries of the ring as must the eigenvectors on the right side, but that is accomplished in a very different way.  They exhibit a pattern which is twice repeated in going around the ring once unlike the eigenvectors right of the transition, which just do not involve half the sites (those initially unoccupied).  In addition, there are ``level'' crossings where the association between eigenvectors and eigenvalues switch back and forth, and thus there there is the possibility  of transitions in the dynamics with regards to which subspaces dominate the production of saddles.  On a final note, if one chooses an initial coherent state with all of its particles in a single site, there is a generally similar appearance to the spectral dependence on $\theta$.  There does not appear to be qualitatively new dynamical  features associated with initially occupying a single site relative to the density wave example.

There are initial conditions though which do.  For example, it is straightforward to show using the mean field (Hamiltonian) equations of motion that the trajectory associated with equal site populations and phases of $b$ is stable for all values of $J,U$.  Its spectrum must behave quite differently than the density wave.  Consider a $4$-site ring populated $|20,20,20,20\rangle$ with all $b=\sqrt{20}$.  Figure \ref{fig8} shows its spectrum in the upper panel and illustrates how different the behavior can be from the coherent state density wave example.  It turns out that the largest eigenvalue is associated with the eigenvector perpendicular to the energy surface, which just represents the associated dynamical shearing.  One of the eigenvalues is doubly degenerate and only $3$ curves are apparent.  A 
\begin{figure}[tbh]
\includegraphics[width=8.5 cm]{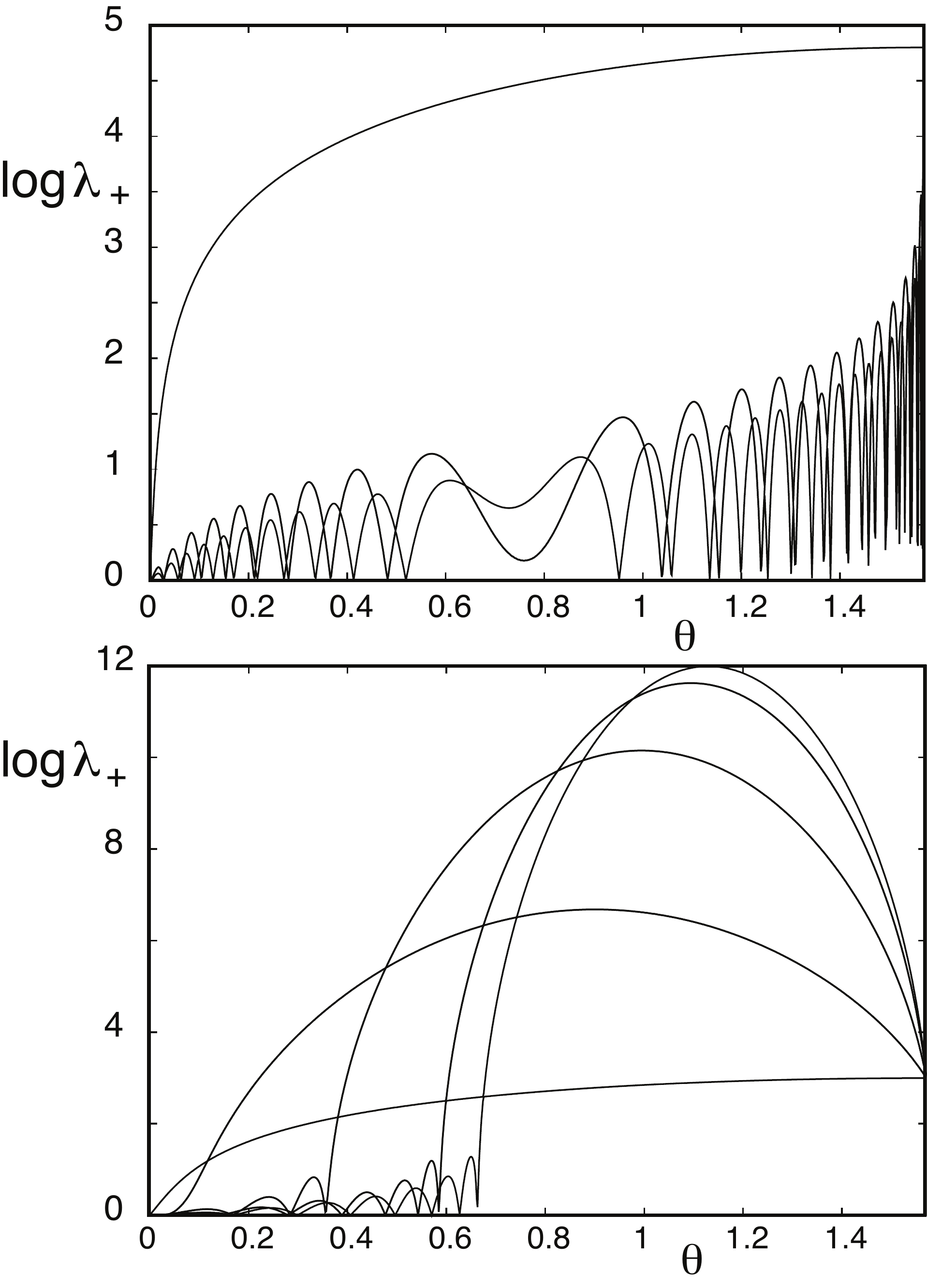}
\caption{Spectrum, $\{\log_{10}\left(\lambda_{j,+}\right)  \}$, of ${\bf M}_\tau \cdot {\bf A}_\alpha^{-1} \cdot {\bf M}_\tau^T$ as a function of $\theta$. The upper panel shows the $\{\log_{10}\left(\lambda_{j,+}\right)  \}$ of a  $4$-site ring with equal populations and phase relations; see text for details.  The lower panel shows the $\{\log_{10}\left(\lambda_{j,+}\right)  \}$ for an $8$-site ring with an alternating phase relationship from site to site; see text for details. There is no abrupt dynamical transition for these examples as there is for coherent state density waves.  There is a transition to chaotic dynamics in the lower panel, and the greatest degree of instability seen for any trajectories. \label{fig9}}
\end{figure}
small change to this coherent state, i.e.~alternating the sign of the $b_j$ creates the most unstable dynamics that we have seen in calculations.  This is such a strong effect that the mean site particle number had to be reduced to $2.5$ to prevent the instability from exceeding the precision available in the calculation.  The example in the lower panel is for $8$-sites with $b=(-1)^{j+1}\sqrt{2.5}$ for the $j^{th}$ site.  There is no abrupt transition for this initial state, but there are a number of level crossings where the dominant dynamical features are interchanged.  There appears to be only $5$ eigenvalues because $3$ of them are doubly degenerate.  They are the ones which on the right side of the figure are interior to the highest and lowest eigenvalues.  This is also where the lowest eigenvalue is the one associated with the normal to the energy surface and the most unstable to the creation of doubly degenerate saddles.

\section{Summary}

Gaussian wave packets and their intimately related counterparts, coherent states for bosonic many-body systems, have great importance in a wide variety of fields.  With respect to their dynamics in systems far from equilibrium, i.e.~short wavelength or mesoscopic regimes, semiclassical methods are ideally suited to furnish excellent quantitative approximations and physical pictures of the essential physics.  Nevertheless, they have rarely been applied completely to wave packet dynamics for systems with more than a couple of degrees of freedom.  The dual problems of performing complex trajectory saddle point searches with many parameters, and determining which ones must be kept due to Stokes phenomena present formidable barriers to the development of practical techniques for implementing the theory fully.   

In this paper, a technique similar in spirit to the tangent space decomposition method for calculating Lyapunov exponents~\cite{Gaspard98,Ott02} and the anisotropic method~\cite{Sala16} is developed to identify the minimal search space.  From the minimal space, the method only relies on identifying real transport pathways and a Newton-Raphson scheme introduced earlier~\cite{Pal16}.  Any system, independent of its number of degrees of freedom, with a small number of dominant expansion directions can be treated.  With these techniques, it has been demonstrated that thousands of saddles can be located in individual systems possessing up to $8$ degrees of freedom.  That particular Bose-Hubbard model case requires a minimal $3$-dimensional parameter search space; the high symmetry, low multiplicity saddles require even smaller dimensional searches.  On the other hand, a straightforward search without the stability analysis would have required a $16$-dimensional parameter search space.  That would have rendered the saddle search effectively impossible to carry out.  Up to the propagation times considered, the set of saddles identified is essentially complete, which can be partly confirmed by comparing a Monte Carlo method applied to the classical transport with the diagonal approximation of the semiclassical quantities.  Furthermore, with a complete knowledge of the saddles, it was shown in~\cite{Tomsovic18b} that a semiclassical theory could capture post-Ehrenfest interference phenomena in the context of the Bose-Hubbard model in a ring configuration extremely accurately.

The existence of symmetries in the system dynamics imposes a significant structure in the locations and multiplicities of symmetry related saddle trajectories, depending on the choice of system state being propagated.  Understanding the fundamental domains, which follows from the group operations involved and the eigenvectors of ${\bf A}^{-1}_\alpha (\tau)$ multiplied by ${\bf M}_\tau^{-1}$, allows one to reduce the search space further.  Symmetry also has a strong influence on the dynamics.  It turns out that high symmetry, low multiplicity saddles dominate the earliest return dynamics that later give way to dominance by low symmetry, high multiplicity saddles.  The high multiplicity saddles generate constructive interference, and enhance long time averages of quantities such as the autocorrelation function.  For the far-out-of-equilibrium dynamics of a many-body system such as the Bose-Hubbard model discussed, symmetry related saddles necessarily lead to constructive interference, and any enhancement factor is revealed over time, not immediately, depending on the time scales at which the various saddle multiplicities are dominant.   An example was shown of the time dependence of the enhancement factors.  There the transition of the enhancement factor from $1\rightarrow 3 \rightarrow 6$ occurred over just a few Ehrenfest times, but for other cases, such as the $8$-site case mentioned, and in other dynamical regimes, it can take much longer for the full enhancement to settle into the dynamics.  All of this information is captured in a full semiclassical theory incorporating quantum interference through the properties of the saddles.

The dynamical analysis relying on the spectrum of ${\bf A}^{-1}_\alpha (\tau)$ and the associated eigenvectors of initial conditions found with the application of ${\bf M}_\tau^{-1}$ can be turned into a powerful and quick way to investigate various dynamical regimes and possibilities of multidimensional dynamical systems, especially for those depending on tunable parameters.  As illustrated with the Bose-Hubbard model, high degrees of instability or abrupt dynamical transitions are easily identified.  Commonalities also appear evident, such as the similarities seen on varying site numbers or scaling with particle numbers.   The eigenvectors also must reflect the symmetries of the system, but there may be multiple ways of accommodating them in high dimensional spaces.  Any transitions between such regimes are associated with spectral crossings that indicate where they occur in the parameter space.

 Building on the work here, there are a large number of directions that future research could go.  There are many other kinds of quantities of interest that can be pursued.  There are other classes of states, such as Fock states, that would require modifying the implementation techniques.  In addition, there are entanglement measures, and out-of-time-ordered correlators, questions regarding thermalization, and relaxation in many-body systems that would be of interest as well.  There are also spectroscopic problems that could be addressed, such as found in molecular spectroscopy, femtosecond chemistry, or attosecond physics.  The beginning would be to identify the equivalent Lagrangian manifolds associated with the quantities of interest, and adapting the search methods to the relevant manifolds.

\begin{appendix}

\section{Associating coherent state and wave packet parameter sets}
\label{cswp}

First consider the usual quantum harmonic oscillator in $1$ degree of freedom,
\begin{equation}
H(\hat p,\hat x) = \frac{\hat p^2}{2m} + \frac{m\omega^2}{2} \hat x^2
\end{equation}
with the creation operator
\begin{equation}
\hat a^\dagger = \frac{1}{\sqrt{2\hbar}}\left( \sqrt{m\omega}\hat x - i \frac{\hat p}{\sqrt{m\omega}} \right)
\end{equation}
The projection of a coherent state into a configuration space representation follows as
\begin{eqnarray}
\label{csx}
\langle x | z \rangle &=& \langle x| \exp \left(-\frac{\left| z \right|^2}{2} + z \hat a^\dagger \right)| 0\rangle \nonumber \\
&=& \exp \left(-\frac{\left| z \right|^2}{2}\right) \sum_{n=0}^\infty \frac{ z^n}{\sqrt{n!}} \langle x | n \rangle \nonumber \\
&=& \left(\frac{m\omega}{\pi\hbar}\right)^{\frac{1}{4}}{\rm e}^{-\frac{\left| z \right|^2}{2} -\frac{m\omega x^2}{2\hbar}} \sum_{n=0}^\infty \frac{ z^n}{\sqrt{2^n}n!} H_n\left(\sqrt{\frac{m\omega}{\hbar}}x\right)\nonumber \\
&=& \left(\frac{m\omega}{\pi\hbar}\right)^{\frac{1}{4}} \exp \left(-\frac{\left| z \right|^2}{2} -\frac{z^2}{2} -\frac{m\omega x^2}{2\hbar} +\sqrt{\frac{2m\omega}{\hbar}} xz\right) \nonumber \\
\end{eqnarray}
where the $H_n(x)$ are Hermite polynomials, and last line follows from an application of the definition of their generating function.  Therefore, the application of the exponential of the creation operator is just a configuration space shift of the ground state multiplied by a global phase.

The following associations of parameters puts the wave packet and position representation of the coherent state into the same form.  Let
\begin{equation}
\label{assoc}
 m\omega = b_\alpha \  {\rm and }\ z=\sqrt{\frac{b_\alpha}{2\hbar}}\left(q_\alpha +i \frac{p_\alpha}{b_\alpha}\right)
\end{equation}
then the configuration space representation of the coherent state is
\begin{eqnarray}
\label{csx2}
\langle x | z \rangle &=&  \exp\left( -\frac{b_\alpha}{2\hbar} (x-q_\alpha)^2 + \frac{i}{\hbar}p_\alpha(x-q_\alpha) +\frac{i}{2\hbar}p_\alpha q_\alpha  \right) \nonumber \\
&& \left(\frac{b_\alpha}{\pi\hbar}\right)^{\frac{1}{4}}
\end{eqnarray}
which is to be compared to Eq.~(\ref{wavepacket}) reduced to its $1$ degree of freedom form,
\begin{equation}
\phi_\alpha(x) = \left(\frac{b_\alpha}{\pi\hbar}\right)^{\frac{1}{4}} \exp\left[ - \frac{b_\alpha}{2\hbar}\left(x - q_\alpha\right)^2+\frac{i}{\hbar}  p_\alpha  \left( x - q_\alpha \right)\right]  
\end{equation}
With the parameter association of Eq.~(\ref{assoc}), the only distinction between the two states is the phase convention.  The wave packet form does not include the phase $\exp[ip_\alpha q_\alpha/(2\hbar)]$, which is easily taken into account.

Next consider an $N$-degree-of-freedom set of coupled harmonic oscillators,
\begin{equation}
H({\bf \hat p},{\bf \hat x}) = \frac{{\bf \hat p} \cdot {\bf \hat p} }{2m} + \frac{m}{2} {\bf \hat x} \cdot {\bf A} \cdot {\bf \hat x}
\end{equation}
As throughout the entire paper, implicitly the right vectors are column vectors and the left vectors are row vectors.  There is an orthogonal transformation to normal coordinates for the column vectors, 
\begin{equation}
{\bf \hat x^\prime} = {\bf O}\cdot {\bf \hat x} \ \  {\rm and}\ \ {\bf \hat p^\prime} = {\bf O}\cdot {\bf \hat p}
\end{equation}
such that 
\begin{equation}
H({\bf \hat p^\prime},{\bf \hat x^\prime}) = \frac{{\bf \hat p^\prime} \cdot {\bf \hat p^\prime} }{2m} + \frac{m}{2} {\bf \hat x^\prime} \cdot {\bf \Omega^2} \cdot {\bf \hat x^\prime}
\end{equation}
where ${\bf \Omega}$ is the diagonal matrix
\begin{equation}
{\bf \Omega} =   \left( \begin{matrix} 
      \omega_1 & 0 & 0 & \\
      0 & \omega_2 & 0 & \hdots\\
     0 & 0 & \omega_3 & \\
       & \vdots &  & \ddots\\
    \end{matrix}\right)
\end{equation}
and
\begin{equation}
{\bf \Omega^2} = {\bf O}\cdot {\bf A} \cdot {\bf O}^T
\end{equation}
The ground state in normal coordinates is
\begin{equation}
\langle {\bf \hat x^\prime} |{\bf 0}\rangle = \left(\frac{m^N {\rm Det}(\Omega)}{\pi^N\hbar^N}\right)^{1/4}\exp \left(  -\frac{m}{2\hbar} {\bf \hat x^\prime} \cdot {\bf \Omega} \cdot {\bf \hat x^\prime} \right)
\end{equation}
As ${\bf A}$ is symmetric and positive definite, it can be decomposed as ${\bf A = B^T \cdot B}$ (with ${\bf B = \Omega\cdot O}$).  Thus, the ground state can also be written in the original coordinates as
\begin{equation}
\langle {\bf \hat x} |{\bf 0}\rangle = \left(\frac{m^N {\rm Det}(\Omega)}{\pi^N\hbar^N}\right)^{1/4}\exp \left(  -\frac{m}{2\hbar} {\bf \hat x}\cdot {\bf B^T} \cdot {\bf \Omega}^{-1} \cdot {\bf B} \cdot {\bf \hat x} \right)
\end{equation}
From this equation, it is already clear that 
\begin{equation}
\label{assocn}
{\bf b}_\alpha = m {\bf B^T} \cdot {\bf \Omega}^{-1} \cdot {\bf B}
\end{equation}
since the action of the exponential of the creation operators is a displacement of the ground state not a deformation.

Assume the coherent state is defined in terms of the creation operators associated with the original coordinate system.  Further, let's project it to begin with onto the normal coordinates.  Thus, the initial quantity to evaluate is
\begin{equation}
\langle {\bf \hat x^\prime} | {\bf z} \rangle = \langle {\bf \hat x^\prime} | \exp \left(-\frac{{\bf z}\cdot {\bf z}^\dagger}{2} + {\bf z}\cdot {\bf  \hat a^\dagger} \right)| {\bf 0} \rangle
\end{equation}
Transforming the creation operators to those associated with the normal coordinates leads to the identifications, $({\bf  \hat a^\dagger})^\prime = {\bf O}\cdot {\bf  \hat a^\dagger}$ for the column vector and   ${\bf z}^\prime =  {\bf z} \cdot {\bf O}^T$ for the row vector.  At this point, action of the exponential of the $({\bf  \hat a^\dagger})^\prime$ is just $N$ independent translations.  This gives, 
\begin{equation}
\label{assoc1}
{\bf z}^\prime = \sqrt{\frac{m {\bf \Omega}}{2\hbar}}\cdot \left({\bf q}^\prime_{\alpha} +i (m{\bf \Omega})^{-1}\cdot{\bf p^\prime}_{\alpha}\right)
\end{equation}
or in component form
\begin{equation}
\label{assoc3}
z_j^\prime = \sqrt{\frac{m \omega_j}{2\hbar}} \left(q^\prime_{\alpha,j} +i \frac{p_{\alpha,j}^\prime}{m\omega_j}\right)
\end{equation}
and that implies for the column vectors of the translations
\begin{eqnarray}
{\bf q}_\alpha &=& {\bf O^T} \cdot {\bf q^\prime}_\alpha \nonumber \\
{\bf p}_\alpha &=& {\bf O^T} \cdot {\bf p^\prime}_\alpha 
\end{eqnarray}
which along with Eq.~(\ref{assocn}) associates the multidimensional coherent state parameters with the wave packet parameters, except for a global phase convention, which is not of great interest.

Returning to just $1$ degree of freedom, there is the possibility of introducing chirps in wave packets as mentioned in the text, which corresponds to the introduction of a complex width $b_\alpha$.  It is well known that free particle motion introduces a complex width parameter as a function of time.  The classical essence of this effect is a linear canonical transformation of the shearing taking place in the dynamics.  As a linear canonical transformation can be associated with an exact unitary transformation in quantum mechanics, a natural way to introduce this effect into a coherent state is to consider the ground state of the Hamiltonian,
\begin{eqnarray}
\label{px}
H(\hat p,\hat x) &=& \frac{(\hat p+\epsilon m \omega x)^2}{2m} + \frac{m\omega^2}{2} \hat x^2 \nonumber \\
&=& \frac{\hat p^2}{2m} +\frac{\epsilon \omega}{2}\left( \hat x \hat p + \hat p \hat x \right) + \frac{1+\epsilon^2}{2} m\omega^2 \hat x^2m\nonumber \\
\end{eqnarray}
The ground state energy remains $E_0=\hbar \omega/2$ and the eigenfunction a Gaussian, but the width becomes complex and it turns out that $b_\alpha = m\omega(1+i\epsilon)$.  A phase convention can be absorbed into the expression for the normalization.  A coherent state can be defined in exactly the same way as in Eq.~(\ref{csx}) with suitably transformed annihilation and creation operators possessing the same properties. The only change is the complexification of $b_\alpha$, which shows up in that equation with the replacement of $m\omega$ with $m \omega (1+i\epsilon)$ with the exception of the normalization factor, which is unchanged.  Thus, $b_\alpha$ is replaced with $(b_\alpha +b_\alpha^*)/2$ in the normalization.  The form of Eq.~(\ref{csx2}) emerges again, only with a complex $b_\alpha$, except that the global phase factor is more complicated.

\end{appendix}

\section*{Acknowledgments}

The author gratefully acknowledges a very helpful critical reading of an early draft of the manuscript by D.~Ullmo and important discussions with D.~Ullmo, P.~Schlagheck, J.~D.~Urbina, K.~Richter, and L.~Kocia.  The author also gratefully acknowledges support from the Vielberth Foundation and the UR International Presidential Visiting Fellowship 2016 during two extended stays at the Physics Department of Regensburg University.

\bibliography{quantumchaos,general_ref,manybody,molecular,classicalchaos}

\end{document}